# High Pressure Effects

James S. Schilling
*Department of Physics, Washington University*
CB 1105, *One Brookings Dr., St. Louis, Missouri 63130*

March 28, 2006

**Abstract**

Experiments under hydrostatic and uniaxial pressure serve not only as a guide in the synthesis of materials with superior superconducting properties but also allow a quantitative test of theoretical models. In this chapter the pressure dependence of the superconducting properties of elemental, binary, and multi-atom superconductors are explored, with an emphasis on those exhibiting relatively high values of the transition temperature $T_c$. In contrast to the vast majority of superconductors, where $T_c$ decreases under pressure, in the cuprate oxides $T_c$ normally increases. Uniaxial pressure studies give evidence that this increase arises mainly from the reduction in the area $A$ of the $CuO_2$ planes ($T_c \propto A^{-2}$), rather than in the separation between the planes, thus supporting theoretical models which attribute the superconductivity primarily to intraplanar pairing interactions. More detailed information would be provided by future experiments in which the hydrostatic and uniaxial pressure dependences of several basic parameters, such as $T_c$, the superconducting gap, the pseudo-gap, the carrier concentration, and the exchange interaction are determined for a given material over the full range of doping.

# 1 Introduction

Pressure, like temperature, is a basic thermodynamic variable which can be applied in experiment over an enormous range, leading to important contributions in such diverse areas of science and technology as astrophysics, geophysics, condensed matter physics, chemistry, biology, and food processing [1, 2]. The field of superconductivity is no exception. The first high-pressure studies on a superconductor were carried out in 1925 by Sizoo and Onnes [3] and revealed that for Sn and In, as for most superconductors [4], the superconducting transition temperature $T_c$ *decreases* under pressure. As will be seen below, the explanation for this pressure-induced decrease in $T_c$ rivals the isotope effect in its simplicity.

It is no accident that many groups active in the synthesis of novel superconducting materials, particularly the many outstanding scientists who emerged from the groups of Bernd Matthias in La Jolla or Werner Buckel in Karlsruhe, routinely use the high-pressure technique as an important diagnostic tool. Why? Because high pressure experiments can provide valuable assistance in the search for superconductors with higher values of $T_c$. In contrast to magnetic materials, which owe their enormous technological importance to the fact that their magnetism is stable to temperatures well above ambient, current materials do not become superconducting unless they are artifically cooled to temperatures at least 160 K below ambient, an inconvenient and expensive process in large-scale applications. An overriding goal in technology-oriented superconductivity research is, therefore, to find materials where $T_c$ surpasses room temperature.

One way to estimate whether a new superconducting material is capable of higher $T_c$ values is to determine how much $T_c$ changes under variation of the chemical composition and/or the pressure. A large value of $|dT_c/dP|$ gives hope that higher values of $T_c$ are possible. We give three examples. A notably successful application of this strategy were the early high-pressure experiments of Paul Chu's group [5] on the $La_{2-x}Ba_xCuO_4$ cuprate (La-214); the large value of $dT_c/dP$ (+8 K $GPa^{-1}$) led to the substitution of the smaller $Y^{3+}$ cation for $La^{3+}$ and the discovery of the famous $YBa_2Cu_3O_{7-\delta}$ compound (Y-123), the first superconductor with $T_c$ above the boiling point of liquid $N_2$ ($\sim$ 77 K). A 2nd example: in the oxide $La_{2-x}Sr_xCuO_4$ $T_c$ is found to increase if compressed in one direction, but decrease if compressed in another [6]; Locquet *et al.* [7] used this fact to appropriately strain thin films of this oxide by growing them epitaxially on a subtrate, thus doubling the value of $T_c$ from 25 K to 49 K. In a 3rd example, the observed increase in $T_c$ under pressure for the Hg-oxides [8] (and for most cuprate oxides for that matter) prompted very high pressure experiments on $HgBa_2Ca_2Cu_3O_8$ (Hg-1223) whereby $T_c$ increased from 134 K to temperatures near 160 K [9]. In less than 10 years, therefore, the record high value of $T_c$ increased *sevenfold* from 23 K for $Nb_3Ge$ to $\sim$ 160 K for Hg-1223. A further increase by only a factor of two would place $T_c$ above room temperature! Forty years ago Neil Ashcroft [10] raised the possibility that elemental hydrogen may become a room temperature superconductor, if only sufficient pressure is applied. The metallization of hydrogen and the development of a viable theory for high-$T_c$ cuprates remain two central goals in current condensed matter research.

High-pressure experiments contribute to the field of superconductivity in diverse ways: (1) As mentioned above, a large magnitude of $dT_c/dP$ is a good indicator that higher values of $T_c$ may be possible for a given superconductor at ambient pressure through chemical substitution or epitaxial growth techniques. (2) Some superconducting materials can only be properly synthesized through the simultaneous application of high pressure and high temperature [11]. (3) Many nonsuperconducting materials become superconducting if sufficiently high pressures are applied. As seen in Fig. 1, there are 29 elemental superconductors at ambient pressure. Under pressure 23 more become superconducting (Li, B, O, Si, P, S, Ca, Sc, Fe, Ge, As, Se, Br, Sr, Y, Sb, Te, I, Cs, Ba, Bi, Ce, Lu); almost half of these were discovered by Jörg Wittig in the 1960's and 1970's [1, 12]. (4) The basic electronic and lattice properties of a material change with decreasing temperature due to the thermal contraction of the lattice. High pressure experiments change the lattice parameters directly at any temperature and thus allow one to correct for the thermal contraction effects at ambient pressure, yielding isochores. (5) Determining the dependence of $T_c$ and other superconducting properties on the individual lattice parameters of a single sample allows a clean quantitative test of theoretical models and gives information on the pairing mechanism. For example, if superconductivity in the high-$T_c$ cuprates results primarily from interlayer coupling, one would anticipate a particularly strong change in $T_c$ if uniaxial pressure is applied perpendicular to the layers.

Unfortunately, all high pressure experiments are not created equal! In superconductivity the pressure dependence of $T_c$ may depend on the pressure medium used and other factors, as illustrated in Fig. 2 for Pb: $T_c(P)$ using the relatively stiff pressure medium methanol:ethanol lies clearly above that when helium is used [15]. Ideally, the applied pressure should be either purely hydrostatic or purely uniaxial. A purely hydrostatic experiment, however, is only possible over a limited pressure/temperature range since all fluids solidify under pressure, the last one being liquid He which requires 12 GPa to freeze at room temperature. Solid He is very soft, i.e. it can only support very weak shear stresses. Dense He is, therefore, the pressure medium of choice in high pressure experiments [16, 17]. One practical way to test whether or not a given experimental result is sensitive to shear stress effects is to carry out the experiment using *two different* pressure media; if the pressure dependence in question remains the same, it is unlikely that shear stress effects play a major role.

This Chapter will restrict itself primarily to the final (5th) benefit of high-pressure investigations as applied to elemental, binary, and multi-atom superconductors. The focus will be on those materials with the highest values of $T_c$ since it can be argued that a thorough understanding of such materials will be most likely to lead to further increases in $T_c$. Attaining the highest values of $T_c$ demands careful optimization of the relevant electronic and lattice (structural) properties. This optimization is most difficult to realize in elemental solids; here the maximum value of $T_c$ has been limited to the temperature range 9 - 20 K (for Nb at ambient pressure and for Li, B, P, S, Ca, V, Y, Zr, and La under very high pressures, as seen in Fig. 1). It is not surprising that multi-atom systems exhibit higher values of $T_c$ since their structural flexibility allows a higher degree of optimization. The highest values of $T_c$ are exhibited by quasi 2D solids such as $MgB_2$ and the high-$T_c$ cuprate oxides. The cuprates, however, exhibit great structural and electronic complexity under both ambient and high pressure conditions, a fact which has greatly hampered attempts to reach a basic understanding of the physical mechanisms responsible for the superconducting state. We will, therefore, begin by discussing in some detail what we can learn from high pressure experiments on the relatively simple elemental and binary superconductors before tackling the much more difficult high-$T_c$ oxides.

Rather than attempting to review the results of high pressure studies on all known superconducting materials, this Chapter will attempt to highlight the *new* information high-pressure experiments provide, information not readily available using other techniques. We refer the reader to excellent reviews of the relatively low-$T_c$ heavy Fermion [19, 20, 21, 22, 23] and organic [24] superconductors which are not included here.

# 2 Elemental Superconductors

Referring to Fig. 1, let us begin by considering those superconductors where the pressure dependence of $T_c$ is easily understood, namely, the ten simple $s, p$ metals which are superconducting at ambient pressure: Be, Al, Zn, Ga, Cd, In, Sn, Hg, Tl, Pb. Under sufficiently high pressures, the number of simple $s, p$ metal superconductors is increased by fourteen: Li, B, O, Si, P, S, Ge, As, Se, Br, Sb, Te, I, Bi. The four $s, p$ elements Cs, Ca, Sr, and Ba also become superconducting under pressure, but their superconductivity is likely rooted in the fact that they exhibit strong $s \rightarrow d$ transfer under pressure and thus effectively become transition metals. The remaining 24 elemental superconductors in Fig. 1 are either transition metals, rare-earth metals, or actinide metals for all of which the conduction electron character is dominated by $d$-electrons.

## 2.1 Simple Metals

### 2.1.1 Non Alkali Metals

The isotope effect played a pivotal role in the development of the BCS theory [25] for conventional phonon-mediated superconductivity. This is due to the fact that isotopic substitution primarily affects only a *single* property, the lattice vibration (phonon) spectrum. Considering the BCS expression in Eq (1), changes in the the isotopic mass $M$ primarily affect the *prefactor* $\Theta_D$, the Debye temperature, and not the exponent, whence the simple relation $T_c \propto \Theta_D \propto M^{-\frac{1}{2}}$. On the other hand, if a superconductor is subjected to high pressures, the *exponent* in Eq (1) is affected since important changes in *both* the lattice vibration *and* the conduction electron states occur. The dependence of $T_c$ on pressure, therefore, may be rather complex, as we shall see below. However, in simple $s, p$ metal superconductors like Sn, In, Zn, Pb, and Al, the pressure-induced stiffening of the lattice vibration spectrum completely dominates over the minor changes in the electronic properties, leading to a *ubiquitous decrease* in $T_c$ with increasing pressure [26], as seen, for example, in Fig. 2 for Pb. Pressure-induced structural phase transitions in simple metals may prompt $T_c$ to jump to higher (or lower) values [27], but otherwise $dT_c/dP$ exhibits a negative slope. A diamond-anvil cell was first used to study superconductivity in the beautiful experiments by Gubser and Webb [28] on Al in 1975; $T_c$ was found to decrease under $\sim$ 6 GPa pressure *fifteenfold* from 1.18 K to 0.08 K.

The above discussion can be made more concrete by analyzing the BCS expression [25] for the transition temperature

$$T_c \simeq 1.13\Theta_D \exp\left\{\frac{-1}{N(E_f)\mathcal{V}_{eff}}\right\}, \tag{1}$$

where $N(E_f)$ is the electronic density of states at the Fermi energy and $\mathcal{V}_{eff}$ is the effective attractive pairing interaction (for simplicity we set $k_B = \hbar = 1$). Since the $s, p$ electrons in simple metals are normally nearly free, one expects in a 3D system $N(E_f) \propto V^{+2/3}$ so that under pressure $N(E_f)$ decreases even more slowly than the sample volume $V$. The principal source for the observed decrease in $T_c$ with pressure in simple $s, p$ metals, however, is not the decrease in $N(E_f)$ but the sizeable decrease in the pairing interaction $\mathcal{V}_{eff}$ itself. The argument can be made more explicit by neglecting the Coulomb repulsion and using McMillan's [29] expression for the electron-phonon coupling parameter $\lambda$

$$N(E_f)\mathcal{V}_{eff} = \lambda = \frac{N(E_f)\langle I^2\rangle}{M\langle\omega^2\rangle}, \tag{2}$$

where $\langle I^2\rangle$ is the average square electronic matrix element and $\langle\omega^2\rangle$ the average square phonon frequency. Making the simplifying assumptions that $\Theta_D \approx \langle\omega\rangle \approx \sqrt{k/M}$, where $k$ is the lattice spring constant, and $M\langle\omega^2\rangle \approx M\langle\omega\rangle^2 \approx M(k/M) = k$, Eq (1) becomes

$$T_c \approx \sqrt{\frac{k}{M}} \exp\left\{\frac{-k}{N(E_f)\langle I^2\rangle}\right\}. \tag{3}$$

In the isotope effect, $M$ appears explicitly only in the prefactor, so that one obtains the canonical BCS relation $T_c \propto M^{-\frac{1}{2}}$. In a high pressure experiment, the changes in $T_c$ are relatively large since they arise principally from the terms in the exponent. In simple metal superconductors, for example, the quantity in Eq (3) which changes most rapidly under pressure is the spring constant $k$, the denominator in the exponent being only weakly pressure dependent, as we discuss below. As $k$ increases with pressure, the modest increase of the prefactor $\sqrt{k}$ is overwhelmed by the decrease from the $-k$ in the exponent, leading to the universal rapid decrease in $T_c$ with pressure for simple $s, p$ metal superconductors. For example, Al, Sn, and Pb, where $T_c(0) \simeq 1.14$ K, 3.73 K, and 7.19 K, respectively, have the pressure dependences $d \ln T_c/dP \simeq$ -0.25, -0.13, and -0.051 GPa$^{-1}$ [30, 31]. What is also immediately evident from these data is that $|d \ln T_c/dP|$ is largest when $T_c(0)$ is smallest. It can be easily shown that this inverse correlation follows directly from the fact that $T_c$ depends *exponentially* on the solid-state parameters $N(E_f)\mathcal{V}_{eff}$. To show this, take the logarithm of both sides of Eq (1) and then the derivative with respect to pressure to obtain

$$\frac{d \ln T_c}{dP} = \frac{d \ln \Theta_D}{dP} + \left[\ln \frac{\Theta_D}{T_c}\right] \left[\frac{d \ln N(E_f)\mathcal{V}_{eff}}{dP}\right]. \tag{4}$$

The first term on the right side of this equation is normally small and can be neglected. The quantity in the left square bracket is positive. The sign of $d \ln T_c/dP$, therefore, is determined by that of $[d \ln N(E_f)\mathcal{V}_{eff}/dP]$ which is negative in the present case. Since $\ln[\Theta_D/T_c]$ becomes larger for decreasing $T_c$ (unless $\Theta_D$ decreases substantially) the magnitude of $d \ln T_c/dP$ would be expected to increase for smaller $T_c$, as observed. Note that such an inverse correlation would not be obtained were $T_c$ to only depend on some (high) power of the solid-state parameters.

To put this discussion on a more quantitative basis, we consider the McMillan equation [29]

$$T_c \simeq \frac{\langle\omega\rangle}{1.20} \exp\left\{\frac{-1.04(1+\lambda)}{\lambda - \mu^*(1+0.62\lambda)}\right\}, \tag{5}$$

which goes beyond weak coupling and connects the value of $T_c$ with fundamental parameters such as the mean phonon frequency $\langle\omega\rangle$, the electron-phonon coupling parameter $\lambda$, and the Coulomb repulsion $\mu^*$. Within this framework, it can be shown that the anticipated change in $\mu^*$ with pressure is normally very small and can be neglected [32]; here we set $\mu^*$ equal to the constant value $\mu^* = 0.1$. However, one should be aware that this assumption for $\mu^*$ may not hold in a more rigorous theoretical framework [33] where the electron-electron and electron-phonon coupling effects are treated on the same footing; this framework yielded for the alkali metal Li the estimate $T_c \simeq 0.4$ mK, in contrast to the value $T_c \approx 1$ K from conventional electronic structure calculations [34].

Taking the logarithmic volume derivative of $T_c$ in Eq (5), we obtain the simple relation

$$\frac{d \ln T_c}{d \ln V} = -B\frac{d \ln T_c}{dP} = -\gamma + \Delta\left\{\frac{d \ln \eta}{d \ln V} + 2\gamma\right\}, \tag{6}$$

where $B$ is the bulk modulus, $\gamma \equiv -d \ln\langle\omega\rangle/d \ln V$ the Grüneisen parameter, $\eta \equiv N(E_f)\langle I^2\rangle$ the Hopfield parameter [35], and $\Delta \equiv 1.04\lambda[1+0.38\mu^*]\,[\lambda - \mu^*(1+0.62\lambda)]^{-2}$. Eq (6) has a simple interpretation. The first term on the right, which comes from the prefactor to the exponent in the above McMillan expression for $T_c$, is usually very small relative to the second term. The sign of the pressure derivative $dT_c/dP$, therefore, is determined by the relative magnitude of the two terms in the curly brackets. The first "electronic" term involves the derivative of the Hopfield parameter $\eta \equiv N(E_f)\langle I^2\rangle$ which can be calculated directly in electronic-structure theory [36]. McMillan [29] pointed out that whereas individually $N(E_f)$ and $\langle I^2\rangle$ may fluctuate appreciably, their product $\eta \equiv N(E_f)\langle I^2\rangle$ changes only gradually, i.e. $\eta$ is a well behaved "atomic" property. One would thus anticipate that $\eta$ changes in a relatively well defined manner under pressure, reflecting the character of the electrons near the Fermi energy [35]. An examination of high-pressure data on simple $s, p$ metal superconductors, in fact, reveals that that Eq (6) is obeyed if $\eta$ increases under pressure at the approximate rate $d\ln\eta/d\ln V \approx -1$ [17], a result also obtained from electronic structure calculations [37]. We also note that Chen *et al.* [32] derived for $s, p$ metals the approximate expression $d\ln\eta/d\ln V = -[d\ln N(E_f)/d\ln V] - 2/3$ which yields for a 3D free-electron gas $d\ln\eta/d\ln V = -4/3 \approx -1$ .

Let us now apply Eq (6) to an analysis of $dT_c/dP$ for simple-metal superconductors. The expression in the curly brackets is positive since the lattice term is positive ($2\gamma \approx +3$ to $+5$) and dominates over the negative electronic term $d\ln\eta/d\ln V \approx -1$. Since $\Delta$ is always positive and the first term $-\gamma$ is relatively small, the sign of $dT_c/dP$ must be negative. This accounts for the universal *decrease* of $T_c$ with pressure in simple metals due to lattice stiffening.

Let us now consider a specific example in more detail. In Sn $T_c$ decreases under pressure at the rate $dT_c/dP \simeq$ -0.482 K GPa$^{-1}$ which leads to $d\ln T_c/d\ln V \simeq +7.2$ [30]. Inserting $T_c(0) \simeq 3.73$ K, $\langle\omega\rangle \simeq 110$ K [38], and $\mu^* = 0.1$ into the McMillan equation, we obtain $\lambda \simeq 0.69$ from which follows $\Delta \simeq 2.47$. Inserting these values into Eq (6) and setting $d\ln\eta/d\ln V = -1$, we can solve for the Grüneisen parameter to obtain $\gamma \simeq +2.46$, in reasonable agreement with the experimental value $\gamma \approx +2.1$ [30]. Similar results are obtained for other conventional simple metal BCS superconductors [17]. Hodder [39] used the McMillan formula and the measured pressure-dependent phonon spectrum for Pb to estimate $dT_c/dP \simeq$ -0.36 K GPa$^{-1}$, in good agreement with experimental values [18, 30, 26].

From the above it is clear that the observed ubiquitous decrease in $T_c$ with pressure for simple metals results from a weakening of the electron-phonon coupling $\lambda$ due to the shift of the phonon spectrum to higher frequencies. This weakening of $\lambda$ is also primarily responsible for the almost universal decrease in the electrical resistivity of simple metals under pressure [40].

### 2.1.2 Alkali Metals

Alkali metals are widely believed to be simple, nearly free electron metals *par excellence* where each atom donates a single $s$ electron to the conduction band, resulting in a nearly spherical Fermi surface. No alkali metal is known to be superconducting at ambient pressure. *In lieu* of a structural phase transition, high pressure would not be expected to induce superconductivity in an alkali metal since, as discussed above, pressure weakens the electron-phonon coupling $\lambda$. In fact, conventional wisdom tells us that high pressure should enhance the free electron behavior of a metal since compressing a solid normally broadens bands and narrows energy gaps.

It was thus with some trepidation that Lin and Dunn [41] reported in 1986 that above 20 GPa the lightest alkali metal, Li, exhibits both a *positive* resistivity derivative $d\rho/dP$ *and* some type of phase transition near 5 K, perhaps a superconducting transition. The matter attracted little attention until 1997 when Neaton and Ashcroft [42] argued on general grounds that under extreme compression the electronic properties of Li could become quite complex and non-free-electron-like due to the near overlap of the atomic 1$s$ cores; the anticipated enhancement in the electron-lattice interaction would be expected to lead to low-symmetry crystal structures, possible superconductivity, and an increase in the electrical resistivity. These results corraborated earlier electronic structure calculations [43] which indicated band-narrowing and gap-widening in Li under extreme compression, i.e. drastic deviations from free-electron behavior.

Three years later two groups [44, 45] subjected Li metal to very high pressures and reported superconductivity above 20 GPa, $T_c$ rising to temperatures approaching 20 K at 30 GPa in the resistivity onset [44]. In these three studies on Li, either a solid pressure medium was used [41] or no pressure medium at all [44, 45], the sample coming in direct contact with the ultrahard diamond anvils. To determine whether the reported superconductivity might have resulted from shear stresses on the Li sample, a third group [46] surrounded the sample with liquid helium in a diamond-anvil cell, as seen in Fig. 3, resulting in nearly hydrostatic pressure conditions. These studies confirmed that Li does indeed become superconducting at 5 K for 20 GPa, $T_c$ rising rapidly to 14 K at 30 GPa, as seen in Fig. 4. In addition, the superconducting phase diagram $T_c(P)$ of Li was accurately mapped out to nearly 70 GPa; several structural phase transitions are indicated at 20, 30, 67 and possibly 55 GPa. The pressure-induced structural transitions in Li have been investigated to 50 GPa in X-ray diffraction studies [47] and to 123 GPa in very recent optical spectroscopic studies [48] at variable temperatures; a unifying scheme for the structural transition mechanisms in all alkali metals has been proposed [49]. These two results, (1) that Li becomes superconducting under pressure and (2) that $T_c$ increases rapidly with pressure, are quite remarkable and confirm that at elevated densities the electronic structure of Li deviates markedly from that of a free-electron gas, the anticipated Fermi surface becoming highly non-spherical [50].

Neaton and Ashcroft [51] applied a similar analysis to the next heavier alkali metal, Na, predicting similar results to those for Li, but at higher pressures. To date, no pressure-induced superconductivity has been found above 4 K in Na to 65 GPa or in K to 43.5 GPa (to 35 GPa above 1.5 K) [150], nor in Rb above 0.05 K to 21 GPa [52]. Very recent studies [53] show that the melting temperature of Na actually decreases for pressures above 30 GPa, falling particularly rapidly above 80 GPa in the $fcc$ phase before passing through a minimum near 110 GPa. These results give strong evidence for highly anomalous electronic behavior in Na in the pressure range above 30 GPa and the likelihood of superconductivity, particularly in the $fcc$ phase above 80 GPa. Further $s, p$ metal systems which likely exhibit anomalous electronic behavior include S which becomes metallic for $P \geq 85$ GPa with a superconducting transition temperature as high as 17 K at $\sim$ 200 GPa nonhydrostatic pressure [54] and P where $T_c$ reaches 18 K at 30 GPa [55], as seen in Fig. 5.

The first alkali metal to become superconducting under high pressure is Cs [56, 57]. Unlike Li and Na, Cs possesses an empty $d$-band which lies relatively near the Fermi energy. Since it can be shown on general grounds that Cs's half-filled 6$s$-band moves up under pressure more rapidly than the bottom of the empty 5$d$-band [58], electrons from Cs's 6$s$ band are transfered into the 5$d$ band ($s \rightarrow d$ transfer), so that under sufficient pressure Cs becomes, in effect, a transition metal. Nonmagnetic transition metals with their higher electronic density of states are normally superconducting, as seen in Fig. 1. Wittig has shown that Cs becomes superconducting at temperatures between 0.05 K and 1.5 K for quasihydrostatic pressures 11 - 15 GPa [56, 57], respectively, a pressure range over which a number of structural transitions occur. McMahan [59] has estimated that in Cs the $s \rightarrow d$ transfer is complete for $P \geq 15$ GPa. Considerably higher values of $T_c$ appear possible at higher pressures, in spite of Cs's 40$\times$ higher ionic mass (M = 133) compared to Li (M = 7). We note that the transition metal superconductor La (M = 139) reaches values near $T_c \approx 13$ K at 12 GPa (see Fig. 5).

Similar scenarios, including superconductivity, would be expected to occur for the next lighter alkali metals, Rb and K, where pressure-induced 5$s \rightarrow$ 4$d$ and 4$s \rightarrow$ 3$d$ transfer is estimated to be complete at 53 and 60 GPa, respectively [59]. It thus seems likely that under sufficiently high pressures all alkali metals will become superconducting.

Although the superconducting properties of the alkali metals become highly anomalous under extreme compression, these properties can still be understood within a conventional BCS framework where the electron pairing arises through the electron-phonon interaction, as for the other simple $s, d$ metals, and, in fact, for the transition metal superconductors which we now briefly discuss.

## 2.2 Transition Metals

In transition metals the $d$-electron character of the conduction band leads to an enhanced density of states $N(E_f)$ which favors superconductivity at higher temperatures than in simple $s, p$ metals. Because of their importance in technological applications, transition metal superconductors have received a great deal of attention, particularly in the 1960's and 1970's. The status of high pressure experiments on $d$-band metals and their theoretical interpretation in terms of electron-phonon mediated supercon-ductivity were comprehensively reviewed by Smith [60] and Garland and Bennemann [61], respectively, in the early 1970's. These same analyses were successfully applied to later systematic studies on transition metal alloys [62, 63].

Although in the majority of transition metal superconductors $T_c$ decreases with pressure, in many cases $T_c$ is found to increase. A positive sign of $dT_c/dP$ for $d$-band superconductors may be understood as arising from a much more rapid increase of the Hopfield parameter under pressure ($d\ln\eta/d\ln V \approx -3$ to $-4$ [35, 61, 64]) than in $s,p$-band superconductors ($d\ln\eta/d\ln V \approx -1$). If, in Eq (6), the electronic term $d\ln\eta/d\ln V$ becomes larger in magnitude than the lattice term $2\gamma$, $T_c$ would be expected to *increase* with pressure; this is, in fact, observed in V, La, and Zr, for example [60], and is seen in Fig. 5 for V, La, Y, Lu, and Sc.

Another reason that the pressure dependence $T_c(P)$ may be particularly complex in transition metals is that the number of $d$ electrons in the conduction band increases under pressure due to $s \rightarrow d$ transfer [58], enhancing the possibility of pressure-induced structural transitions or electronic (Lifshitz) transitions; such effects are likely responsible for the unusually complex $T_c(P)$ dependence of LaAg where $T_c(P)$ to 2.5 GPa passes through two maxima and minima [74]. See the review by Lorenz and Chu [75] for examples of electronic transitions.

It is well known that the number of $d$ electrons, $n_d$, is a particularly significant quantity in determining the crystal structure and the electronic properties of transition metal [76, 77], rare-earth [78], and actinide [77] solids. The pressure-induced superconductivity in the pre-transition elements Cs, Ca, Sr, and Ba is likely the result of $s \rightarrow d$ electron transfer.

In Fig. 5 $T_c(P)$ data are compared for the trivalent transition metals La, Y, Sc, and Lu. The very recent nearly hydrostatic [70] and nonhydrostatic [71] studies on Y metal differ substantially from earlier quasihydrostatic work [56, 72] and reveal that $T_c$ increases monotonically from 5 K at 30 GPa to 19.5 K (midpoint) or 20 K (onset) at 115 GPa, the highest value of $T_c$ ever measured in the magnetic susceptibility for an elemental superconductor (see Fig. 1); remarkably, the dependence of $T_c$ on sample volume is nearly linear over the entire pressure range 33 to 115 GPa [70, 71]. The initial slope $dT_c/dP \approx +1$ K GPa$^{-1}$ for La is particularly large, possibly due to the anomalously low value of the Grüneisen constant $\gamma \approx 1$ for this metal [61]. Experiments on V metal show that $T_c$ increases slowly, but nearly linearly, with pressure (+0.1 K GPa$^{-1}$) from 5 K to 17 K at 120 GPa [69]. Unlike for $s,p$ metals, the pressure dependence $T_c(P)$ for transition metals follows no universal behavior, reflecting the additional complexity (and potency!) of the electronic properties in a $d$-electron system.

Can Eq (6) account for the observed pressure dependence of $T_c$ for V? Setting $T_c(0) = 5.3$ K, $\mu^* = 0.1$ and the Debye temperature $\Theta_D = 399$ K [29] in the McMillan equation, where $\langle\omega\rangle = 0.83\Theta_D$, we obtain $\lambda = 0.538$ and thus $\Delta = 3.547$. Inserting now into Eq (6) the volume derivative of the Hopfield parameter ($d\ln\eta/d\ln V \simeq -3.3$) calculated for V by Evans *et al.* [37] and the Grüneisen parameter $\gamma \simeq 1.5$ [79], we obtain $d\ln T_c/d\ln V \simeq -2.56$. Using for the bulk modulus $B = 162$ GPa [79], we obtain, finally, $dT_c/dP = -[T_c(0)/B]d\ln T_c/d\ln V \simeq +0.084$ K GPa$^{-1}$, in good agreement with the experiments of Ishizuka *et al.* (0.1 K GPa$^{-1}$) [69] and the earlier studies of Smith (0.062 K GPa$^{-1}$) [80].

The highest values of $T_c$ yet achieved for an elemental superconductor appear to lie in the range 15-20 K for both *s*-, *p*-, and *d*-electron metals under high pressure. It would be expected that higher values of $T_c$ should be possible for binary or pseudobinary compounds where the flexibility afforded by two elements should allow a superior optimization of the parameters. Indeed, binary superconductors reach values of $T_c$ which are more than twice as high as those for elemental superconductors.

# 3 Binary Superconductors

## 3.1 A-15 Compounds

Until the discovery of the cuprate oxides in late 1986, the binary A-15 compounds $Nb_3Ge$ ($T_c \simeq 23$ K), $Nb_3Sn$ ($T_c \simeq 17.8$ K), and $V_3Si$ ($T_c \simeq 16.6$ K) exhibited the highest values of $T_c$. High pressure studies on the A-15's were reviewed in 1972 by Smith [81]. Hopfield [35] noted that the near doubling of the value of $T_c$ from Nb to $Nb_3Sn$ could be simply understood, using the above relation $d\ln\eta/d\ln V \approx -3.5$, as resulting from an enhancement in $\eta$ by $\sim 60\%$ due to the reduced Nd-Nd separation in $Nb_3Sn$.

In the A-15 compounds the competition between subtle structural transitions and superconductivity has been extensively studied. A case in point are parallel studies by two groups [82, 83] on "nontransforming" $V_3Si$ crystals where $T_c$ increases under pressure from 16.6 K to approximately 17.7 K at 3 GPa, whereupon $T_c(P)$ exhibits a break in slope signalling a cubic-to-tetragonal structural transformation predicted by Larsen and Ruoff [84]. Further details and references on A-15 compounds are contained in a recent review by Lorenz and Chu [75]. We will see below that the high-$T_c$ oxides provide numerous examples for the influence of structural defects and transitions on superconductivity, perhaps more than one would like!

Following the discovery of high-$T_c$ superconductivity in the cuprates, two binary compounds, $MgB_2$ and $Rb_3C_{60}$, were discovered which have substantially higher transition temperatures than the A-15's. We now consider high-pressure studies on these two compounds.

## 3.2 A Special Case: $MgB_2$

The binary superconductor with the highest known value of the transition temperature, $MgB_2$ with $T_c \approx 40$ K, was discovered in early 2001 [85]; Buzea and Yamashita [86] have reviewed its superconducting properties. $MgB_2$ is a quasi-2D material with strong covalent bonding within the graphite-like $B_2$ layers. Understandably, the compressibility is highly anisotropic, being 64% greater along the $c$ axis than the $a$ axis, with bulk modulus $B = 147.2(7)$ [87]. The anisotropy in the superconducting properties is also appreciable, but less than that observed in the high-$T_c$ oxides [88].

Several studies of the dependence of $T_c$ on pressure for polycrystalline $MgB_2$ were carried out shortly after the discovery of its superconductivity [89, 90, 91, 92]. The first studies used either solid (steatite) [89] or fluid (Fluorinert) [90, 91] pressure media and agreed that $T_c$ decreases under pressure, but disagreed widely on the rate of decrease which ranged from -0.35 to -1.9 K GPa$^{-1}$. The first truely hydrostatic measurement of $T_c(P)$ was carried out to 0.7 GPa using He gas on an isotopically pure ($^{11}$B) sample [92]; it was found that $T_c$ decreases reversibly under hydrostatic pressure at the rate $dT_c/dP \simeq -1.11 \pm 0.02$ K GPa$^{-1}$, yielding $d\ln T_c/d\ln V = Bd\ln T_c/dP \simeq +4.16 \pm 0.08$. This latter result was confirmed subsequently by He-gas studies on $MgB_2$ single crystals to 0.6 GPa as well as parallel diamond-anvil-cell studies in dense He to nearly 30 GPa [93] which are shown in Fig. 6; the latter are in excellent agreement to 20 GPa with parallel studies in dense He by Goncharov and Struzhkin [94]. On the other hand, diamond-anvil-cell studies on the same samples using methanol:ethanol [95] or Fluorinert [93] pressure media resulted in a substantially more negative slope $dT_c/dP$, apparently arising from shear stress effects in these frozen pressure media.

Ultrahigh-resolution thermal expansion and specific heat measurements on $MgB_2$ yield through the Ehrenfest relation $dT_c/dP \simeq -1.05 \pm 0.13$ K GPa$^{-1}$, in excellent agreement with the dependence $-1.07 \pm 0.03$ K GPa$^{-1}$ obtained in He-gas studies, all on the same sample [96]. On cooling through $T_c$, both the thermal expansion coefficient and the Grüneisen function change from positive to negative, the latter showing a dramatic increase to large positive values at low temperature. These results suggest anomalous coupling between superconducting electrons and low-energy phonons [96].

We now apply the same analysis carried out above for simple $s, p$ metal superconductors to $MgB_2$ to see whether the measured dependence $dT_c/dP \simeq -1.11 \pm 0.02$ K GPa$^{-1}$ is consistent or not with BCS theory (electron-phonon coupling). Using the average phonon energy from inelastic neutron studies [97] $\langle\omega\rangle = 670$ K, $T_{c0} \simeq 39.25$ K, and $\mu^* = 0.1$, we obtain from the above Eqs (5) and (6) $\lambda \simeq 0.90$ and $\Delta \simeq 1.75$. Our estimate of $\lambda \simeq 0.90$ agrees well with those of other authors [98, 99]. Since the pairing electrons in $MgB_2$ are believed to be $s, p$ in character [100, 101, 98, 102], we set $d\ln\eta/d\ln V = -1$, a value close to that $d\ln\eta/d\ln V = Bd\ln\eta/dP \approx -0.81$, where $d\ln\eta/dP \approx +0.55$ %/GPa, from first-principles electronic structure calculations by Medvedera *et al.* [103]. Inserting the above values of $d\ln T_c/d\ln V = +4.16$, $\Delta = 1.75$, and $d\ln\eta/d\ln V = -1$ into Eq (6), we find for the Grüneisen parameter $\gamma = 2.36$, in reasonable agreement with the values $\gamma \approx 2.9$ from Raman spectroscopy studies [48] or $\gamma \approx 2.3$ from *ab initio* electronic structure calculations on $MgB_2$ [104]. A similar analysis of the data in Fig. 6 to 30 GPa, based on an analysis by Chen *et al.* [32], also gives excellent agreement. See Ref. [93] for a full discussion and a comprehensive summary of all high-pressure studies on $MgB_2$.

The He-gas $T_c(P)$ data are thus clearly consistent with electron-phonon pairing in $MgB_2$, in agreement with high precision isotope effect experiments [105, 106]. The fact that the B isotope effect is fifteen times that for Mg [106] is clear evidence that the superconducting pairing originates within the graphite-like $B_2$ layers.

### 3.3 Doped Fullerenes $A_3C_{60}$

A particularly interesting class of superconductors with high values of $T_c$ are the alkali-doped fullerides $A_3C_{60}$, where A = K, Rb, Cs [107], each alkali atom donating one $s$ electron to the conduction band. $K_3C_{60}$ and $Rb_3C_{60}$ have $T_c$ values of 19 K and 29.5 K, respectively; evidence has been found for superconductivity in $Cs_3C_{60}$ near 40 K [108], but this has yet to be duplicated. The increase in $T_c$ from $K_3C_{60}$ to $Rb_3C_{60}$ to $Cs_3C_{60}$ is mainly related to lattice expansion (negative pressure) effects [109].

$T_c$ for the alkali-doped fullerides is found to *decrease* under the application of hydrostatic pressure [109]. For $Rb_3C_{60}$, for example, $dT_c/dP \simeq -8.7$ K GPa$^{-1}$, as seen in Fig. 7 [110]. Since the bulk modulus of $Rb_3C_{60}$ is given by $B = 18.3$ GPa [111], one can estimate $d\ln T_c/d\ln V = B(d\ln T_c/dP) \simeq +5.4$, a value intermediate between that for $MgB_2$ (+4.16) and Sn (+7.2). It would thus be reasonable to expect that the reason for the negative value of $dT_c/dP$ for the alkali-doped fullerenes is the same as for $MgB_2$, Sn, and other $s, p$ metal superconductors, namely, lattice stiffening.

To test this hypothesis, let's attempt an analysis of the above data in terms of electron-phonon coupling using the above McMillan equation and its pressure derivative, invoking the *intermolecular* lattice vibrations for $Rb_3C_{60}$ which are in the range 15 - 150 K [112]. Setting the average value $\langle\omega\rangle \approx 80$ K and using $\mu^* = 0.2$ [113], Eq (5) yields a *negative* value for $\lambda$, an impossibility, implying that this equation must be invalid for the given set of parameters. Even setting $\langle\omega\rangle \approx 150$ K, the upper limit for intermolecular vibrations, $\lambda \approx 5$ would be required by Eq (5), a value clearly beyond the range of validity of the McMillan equation ($\lambda \leq 1.5$). To proceed, we use the simple expression

$$T_c = \frac{0.26 E_{char}}{\sqrt{e^{2/\lambda} - 1}}, \tag{7}$$

valid for all values of $\lambda$ [114], where $E_{char}$ is the characteristic lattice-vibration energy. Setting $E_{char} = \langle\omega\rangle \approx 80$ K and $T_c(0\text{ K}) = 29.5$ K, Eq (7) yields $\lambda = 5$. Taking the pressure derivative of Eq (7), and using a typical value of the Grüneisen parameter $\gamma \approx +2$, it is easy to show [110] that the above value of $dT_c/dP$ is only possible if $d\ln\eta/d\ln V \approx +10$ ! This value of $d\ln\eta/d\ln V$ differs grossly in both magnitude and sign from that typically found for conventional simple-metal (-1) or transition-metal (-3.5) superconducting elements, alloys, or compounds [35]. What is likely wrong is the above assumption that the *intermolecular* lattice vibrations are responsible for the superconductivity.

On the other hand, if we assume the characteristic lattice-vibration energy is given by the high frequency *intramolecular* (on-ball) vibrational modes, where $E_{char} =$ $\langle\omega\rangle \approx 350$ - 2400 K, then we cannot account for the negative value of $dT_c/dP$ through lattice stiffening since, due to the extreme stiffness of the $C_{60}$ molecule, the average frequency of the on-ball phonons $\langle\omega\rangle$ and the mean square electron-phonon matrix element $\langle I^2\rangle$ are essentially independent of pressure.

So what is responsible for the rapid decrease in $T_c$ under pressure in $Rb_3C_{60}$? Perhaps electronic effects are important here, in contrast to simple $s, p$ electron metals. The answer to this question is provided by measurements of the pressure-dependent electronic density of states $N(E_f)$ which is found [110] to decrease sharply under pressure, as seen in Fig. 7. This decrease is a direct result of the rapid increase in the width of the conduction band as the $C_{60}$ molecules are pressed together.

We are now confronted with a very different situation than in conventional superconductors. Utilizing our knowledge of $N(E_f)(P)$ in the McMillan equation, one can use the pressure-independent value of $\langle\omega\rangle$ as a parameter to obtain the best fit to the experimental $T_c(P)$ data. A detailed analysis [110] reveals that weak-coupling theory can account for the experimental pressure dependences as long as the characteristic energy of the intermediary boson lies between $\langle\omega\rangle \approx 300$ K and 800 K, typical energies for the high frequency on-ball phonons. The reason for the large negative value of $dT_c/dP$ in $Rb_3C_{60}$, therefore, is *not* lattice stiffening, but a sharp decrease in the electronic density of states $N(E_f)$ with pressure. The increase in $T_c$ going from $K_3C_{60}$ to $Rb_3C_{60}$ to $Cs_3C_{60}$ is due mainly to the enhancement in the density of states $N(E_f)$ as the progressively larger interstitial alkali cations expand the lattice, increase the separation between neighboring $C_{60}$ molecules, and thus narrow the conduction band.

# 4 Multi-Atom Superconductors: High-$T_c$ Oxides

As outlined in the Introduction, the high pressure technique led directly to the discovery of $YBa_2Cu_3O_{7-\delta}$ (Y-123) [5], one of the most important high-$T_c$ superconductors (HTSC), and generated in $HgBa_2Ca_2Cu_3O_{8+\delta}$ (Hg-1223) the highest transition temperature $T_c \approx 160$ K [9] for the resistivity onset of any known superconductor (see Fig. 8); very recently Monteverde *et al.* [115] reportedly bested this value by 3-4 K by applying 23 GPa to a fluorinated Hg-1223 sample.

In this section we will attempt to determine the "intrinsic" dependence of $T_c$ on pressure for hole-doped HTSC and from this intrinsic $T_c(P)$ to identify what, if any, new information is provided regarding the mechanism(s) responsible for, and the appropriate theoretical description of, superconductivity in the high-$T_c$ oxides. No attempt will be made to summarize all available results; we refer the reader to previous reviews covering high pressure effects in the high-$T_c$ cuprates: Wijngaarden and Griessen in 1989 [116], Schilling and Klotz in 1991 [17], Takahashi and Môri in 1997 [117], Núñez-Regueiro and Acha in 1997 [118], Lorenz and Chu in 2004 [75], and an all-too-short but interesting paper by Wijngaarden *et al.* in 1999 [119].

We also restrict our consideration here to hole-doped HTSC. As is evident from the above reviews, electron-doped HTSC have received relatively little attention; in the few high-pressure studies carried out, $T_c$ is normally found to decrease with pressure [120]. The fact that electron-doped HTSC must be slightly reduced to induce superconductivity means that oxygen ordering effects will likely play an important role in the pressure dependence of $T_c$, as discussed below for their hole-doped counterparts. Definitive high pressure studies on well characterized electron-doped HTSC which separate "intrinsic" from "oxygen ordering" effects are encouraged.

We begin by showing in Figs. 8 and 9 the pressure dependence of $T_c$ for a number of hole-doped HTSC, including the one-, two- and three-layer Hg-compounds $HgBa_2CuO_{4+\delta}$ (Hg-1201), $HgBa_2CaCu_2O_{6+\delta}$ (Hg-1212), and $HgBa_2Ca_2Cu_3O_{8+\delta}$ (Hg-1223). With the lone exception of $Tl_2Ba_2CuO_{6+y}$ (Tl-1201), $T_c(P)$ is seen to initially *increase* with pressure and pass through a maximum at higher pressures. The nearly ubiquitous initial increase in $T_c$ with pressure, which was first pointed out by Schirber *et al.* [121], is a hallmark of hole-doped high-$T_c$ cuprates.

A central question is whether or not the measured pressure dependence $T_c(P)$ in the superconducting cuprates gives evidence for an unconventional (non electron-phonon) pairing mechanism. As seen in Fig. 5, as for the high-$T_c$ oxides, $T_c(P)$ is known to pass through a maximum for La and S, both of which are believed to superconduct via the standard electron-phonon interaction. As discussed in detail above, for the majority of conventional simple and transition metal superconductors $T_c$ decreases with pressure.

The evident similarity in the $T_c(P)$ dependences for the HTSC systems in Figs. 8 and 9, particularly for the three Hg-compounds, gives strong evidence that the nature of the superconductivity is the same for all. This is not particularly surprising since all HTSC share one common structural element, the $CuO_2$ planes. The question remains, however, whether the most important interactions for the high-$T_c$ superconductivity take place *within* these planes or *between* them. Uniaxial pressure experiments, in particular, hold promise to shed some light on this question.

The pressure dependences $T_c(P)$ in Figs. 8 and 9 bear some resemblence to the canonical inverted parabolic dependence of $T_c(n)$ for HTSC on the hole carrier concentration $n$ per Cu cation in the $CuO_2$ sheet

$$T_c(n) = T_c^{\max}[1 - \beta(n - n_{opt})^2], \tag{8}$$

illustrated in Fig. 10, where $\beta \simeq 82.6$ and $n_{opt} \simeq 0.16$ [123, 124]. According to Eq (8) $T_c(n)$ initially increases with $n$ on the underdoped side from 0 K for $n \approx 0.05$ to a maximum value $T_c^{\max}$ at optimal doping $n = n_{opt}$ before falling back to 0 K for $n \approx 0.27$ on the overdoped side. For underdoped samples one has $dT_c/dn > 0$, for optimally doped $dT_c/dn = 0$, and for overdoped $dT_c/dn < 0$. Since $n$ has been found to initially increase with pressure ($dn/dP > 0$) in the majority of cuprates studied [125, 17, 117, 75], one might conjecture that with increasing pressure $T_c(P)$ simply traces out the inverted parabolic shape of $T_c = T_c(n)$ in Fig. 10, yielding the $T_c(P)$ dependences seen in Figs. 8 and 9. In such a "Simple Charge-Transfer Model", where only $n$ is assumed pressure dependent, the pressure derivative is given by

$$\frac{dT_c}{dP} = \left(\frac{dT_c}{dn}\right)\left(\frac{dn}{dP}\right) = -2\beta T_c^{\max}(n - n_{opt})\frac{dn}{dP}, \quad (9)$$

each system having a particular initial value of $n$. Within this model, the negative value of $dT_c/dP$ for Tl-2201 in Fig. 9 would result from the increase of $n$ with pressure and the well known fact that this compound is overdoped, i.e. $dT_c/dn < 0$.

That "life with the cuprates" is not so simple is seen by the data in Fig. 11(a) on five Y-123 samples for increasing oxygen content $x$ from $A \to E$, four underdoped ($A \to D$) and one nearly optimally doped ($E$). The measured $T_c(P)$ dependences run contrary to the expectations of the "Simple Charge Transfer Model" for underdoped samples, namely, that the higher the initial value of $T_c$, the lower the pressure needed to reach $T_c = T_c^{\max}$. The data on the Hg-compounds in Fig. 8, with initial slope $dT_c/dP \simeq +1.75$ K GPa$^{-1}$, also violate Eq (9). Since all three Hg-compounds are nearly optimally doped, i.e. $dT_c/dn \simeq 0$, one would expect $dT_c/dP \simeq 0$ from Eq (9). Evidently the "Simple Charge Transfer Model" is too simple! Neumeier and Zimmermann [126] extended this model by hypothesizing that the change in $T_c$ with pressure derives from two contributions: (1) an "intrinsic" contribution reflecting pressure-induced changes in $T_c$ resulting solely from the reduction of the lattice parameters (no structural transitions, oxygen ordering effects, nonhydrostatic strains or changes in the carrier concentration) and (2) the above contribution to $dT_c/dP$ in Eq (9) originating from the normal increase in $n$ under pressure. The pressure derivative in this "Modified Charge-Transfer Model" is thus given by the general expression

$$\frac{dT_c}{dP} = \left(\frac{dT_c}{dP}\right)_{intr} + \left(\frac{dT_c}{dn}\right)\left(\frac{dn}{dP}\right). \quad (10)$$

If one now substitutes Eq (8) in this expression and assumes that $\beta$ and $n_{opt}$ are independent of pressure [127], one obtains

$$\frac{dT_c}{dP} = \frac{dT_c^{\max}}{dP}\left[1 - \beta\left(n - n_{opt}\right)^2\right] - \frac{dn}{dP}\left[\left(2\beta T_c^{\max}\right)\left(n - n_{opt}\right)\right]. \quad (11)$$

Note that $T_c^{\max}$ is the maximum value of $T_c$ when the carrier concentration $n$ alone is varied at constant pressure; $T_c^{\max}$ is *not* the maximum value of $T_c$ when the pressure is varied (unless $dn/dP = 0$). Comparing Eqs (10) and (11) we see that the intrinsic component of the pressure derivative is given by

$$(dT_c/dP)_{intr} = (dT_c^{\max}/dP)\left[1 - \beta\left(n - n_{opt}\right)^2\right]. \quad (12)$$

Note that $dT_c/dP = (dT_c/dP)_{intr} \equiv dT_c^{\max}/dP$ only for optimally doped samples, where $n = n_{opt}$. If the sample is nearly optimally doped, then we can neglect the term in Eq (11) quadratic in $(n - n_{opt})$, leaving the following expression linear in $(n - n_{opt})$

$$\frac{dT_c}{dP} = \frac{dT_c^{\max}}{dP} - \frac{dn}{dP}\left[\left(2\beta T_c^{\max}\right)\left(n - n_{opt}\right)\right]. \quad (13)$$

A linear dependence of $dT_c/dP$ on $(n - n_{opt})$ was indeed found in a careful high-pressure (He-gas) study [126] on the $Y_{1-y}Ca_yBa_2Cu_3O_x$ compound series where the Ca and O contents were varied to change $n$ near optimal doping; from the slope of this dependence it was determined that $dn/dP \simeq +0.0055$.holes GPa$^{-1}$. At optimal doping the intrinsic pressure dependence was found to be $dT_c/dP = (dT_c/dP)_{intr} = +0.96$ K GPa$^{-1}$ [126].

We should not be surprised that the "Simple Charge-Transfer Model", which only considers the single charge-transfer contribution, fails to satisfactorily account for the experimental results. We have seen that the $T_c(P)$ dependences for transition metal superconductors can only be understood by taking into account *two* distinct contributions: from both lattice vibrations and electronic properties. One should expect materials as complex as the high-$T_c$ cuprates with their distorted quasi-2D perovskite structures to be a good deal more complex than the transition metals. That this is indeed the case is the reason why it has proven so difficult to reach a basic understanding of HTSC, the results of high-pressure studies being no exception.

Ideally, in a high pressure experiment we would like to determine the change in the superconducting properties of a given high-$T_c$ oxide under variation of both the intraplanar lattice parameter(s) and the interplanar separation. In actual high-pressure experiments, however, a number of additional effects may occur which considerably complicate the interpretation of the data: (1) structural phase transitions, (2) oxygen ordering effects, and (3) effects due to shear stress from non-hydrostatic pressure media.

### 4.0.1 Non-Hydrostatic Pressure Media

As pointed out above, not all high-pressure experiments are created equal. Ideally, a fluid pressure medium is used which transmits hydrostatic pressure to the sample. The use of solid pressure media, or no pressure media at all, may simplify the experimentation, but results in the sample being subjected to varying degrees of non-hydrostatic shear stress which may cause important changes in the superconducting state, in particular in the pressure dependence of the transition temperature $T_c(P)$, as we have seen above for Pb in Fig. 2. Shear-stress effects on $T_c(P)$ are well known from studies on such diverse superconducting materials as organic metals [129], $MgB_2$ [46], Re metal [130], and Hg [131].

The differing $T_c(P)$ results on the high-$T_c$ cuprates by different groups may arise from differences in samples, in the pressure medium, and/or in the method used to determine $T_c$. Gao *et al.* [9] suggested that the fact that their value of $T_c$(30 GPa) $\approx$ 160 K for Hg-1223 lies $10 - 15$ K higher than that found by other groups [117, 132] may have its origin in shear stress effects. Klotz *et al.* [133] carried out two experiments on a single sample of $Bi_2CaSr_2Cu_2O_{8+\delta}$ (Bi-2212) in a diamond-anvil cell, one in helium and the other with no pressure medium whatsoever, and obtained very different $T_c(P)$ dependences. On the other hand, Wijngaarden *et al.* [119]

report that the $T_c(P)$ dependences for $YBa_2Cu_4O_8$ (Y-124) found by different groups using varying pressure media do not differ widely. Also, a recent purely hydrostatic He-gas experiment to 0.6 GPa on an overdoped Y-123 single crystal agrees within experimental error with the initial pressure dependence $dT_c/dP \approx$ -1 K GPa$^{-1}$ found in a parallel diamond-anvil-cell experiment using solid steatite as pressure medium [134]. As discussed in the Introduction, for quantitative investigations fluid pressure media, particularly helium, are to be preferred over solid media. To test whether or not shear stresses play a role in the pressure-induced changes obtained, it is prudent to carry out the experiment using two different pressure media.

#### 4.0.2 Structural Phase Transitions

As for the A-15 compounds, $T_c(P)$ for HTSC can be a sensitive function of structural instabilities. The initial rate of increase of $T_c$ with pressure for $La_{2-x}Sr_xCuO_4$ is relatively large at +3.0 K GPa$^{-1}$ [135]; it is even much larger for $La_{2-x}Ba_xCuO_4$ (+8 K GPa$^{-1}$) [136]. This led Wu *et al.* [5] to the discovery of Y-123, as discussed above. The reason for the anomalously large positive value of $dT_c/dP$ for $La_{2-x}Ba_xCuO_4$ is the existence of a low-temperature-tetragonal (LTT) phase below 60 K which strongly suppresses $T_c$ for $x$ in the range 0.07 to 0.18, as seen in Fig. 12. Applying pressure eventually suppresses this LTT phase transition, leading to the anomalously large increase in $T_c$ under pressure seen in the data where $dT_c/dP$ reaches values as large as +12 K GPa$^{-1}$ [137]. At 2 GPa the $T_c(x)$ dependence in Fig. 12 begins to resemble the canonical bell-shaped $T_c(n)$ dependence of Fig. 10, except in a very narrow range of $x$ centered at $x = 0.125$.

In a further compound system in the same family, $La_{2-x-y}Nd_ySr_xCuO_4$, the doping level or the crystal structure can be independently controlled by varying $x$ or $y$, respectively [138]. In the phase diagram for $La_{1.48}Nd_{0.4}Sr_{0.12}CuO_4$ in Fig. 13 it is seen that at ambient pressure the high-temperature-tetragonal (HTT) phase transforms below 500 K into a low-temperature-orthorhombic (LTO1) phase, followed by a phase change below 70 K to the LTT phase [139]. High pressure is seen to suppress the low-temperature phases until above 4 GPa only the HTT phase remains. These phase transitions are seen in Fig. 13 to have a dramatic effect on the pressure dependence of the superconducting transition temperature $T_c(P)$ which peaks near 5 GPa. Evidently, structural instabilities play an important role in the doped $La_2CuO_4$ oxide family, making it almost impossible to extract the intrinsic dependence of $T_c$ on pressure from experiment.

#### 4.0.3 Oxygen Ordering Effects

In the majority of HTSC oxygen defects are present with a relatively high mobility, even at ambient temperatures. Many HTSC can thus be readily doped simply by varying the oxygen defect concentration through annealing at controlled oxygen partial pressures at elevated temperatures. The normal and superconducting state properties of HTSC depend not only on the concentration of oxygen defects, but on the relative positions assumed by these defects in the lattice on a local scale. Such oxygen ordering effects were first observed at ambient pressure in strongly underdoped Y-123 samples where the $T_c$ value could be sharply reduced simply by quenching the sample from elevated temperatures into liquid nitrogen [140]. A simple model developed by Veal *et al.* [140] was able to account for this phenomenon in terms of a reduction in the hole-carrier concentration $n$ in the $CuO_2$ planes due to reduced local order of oxygen defects in the Y-123 chains containing the ambivalent Cu cations. Oxygen ordering effects on $T_c$ are only *observed* if (1) oxygen defects are present, (2) there are vacant sites available which oxygen defects can move into, and (3) the sample is not optimally doped (if optimally doped, $dT_c/dn = 0$ so that small changes in $n$ due to oxygen ordering have no effect on $T_c$).

A second way to change the oxygen ordering state is through high pressure. The application of high pressure at room temperature prompts the mobile oxygen defects to order locally and thus enhance the hole-carrier concentration $n$ in the $CuO_2$ planes. Pressure-induced oxygen ordering thus "turbo-charges" the normal enhancement of $n$ with pressure. Significant pressure-induced oxygen ordering effects have been observed for Y-123 by Fietz *et al.* [141] and others [128], as illustrated in Fig. 11(right) for an underdoped sample. Whereas the lower $T_c(P)$ curve in this figure was measured in an experiment carried out completely at temperatures low enough ($T < 200$ K) to prevent the ordering of oxygen defects in the chains as the pressure is changed, the upper curve was obtained for pressure changes at ambient temperature. The difference between the two $T_c(P)$ dependences is substantial indeed! In Y-123 the time-dependent relaxation of $T_c$ following a change in pressure can be best fit using the stretched exponent $\beta \simeq 0.6$ [128]. Phillips [142] has argued that this gives evidence for the importance of the electron-phonon interaction in HTSC and supports his model for defect-induced superconductivity [143].

Pressure-induced oxygen ordering effects in HTSC were first observed in overdoped Tl-2201 samples by Sieburger and Schilling [144] and then extensively studied by Klehe *et al.* [145, 146], as illustrated in Fig. 14. If pressure is applied at room temperature, $T_c$ is seen to decrease rapidly, as found earlier by Môri *et al.* [122] (see Fig. 9); however, if the pressure is released at temperatures low enough (55 K) to freeze in the oxygen defects, $T_c$ does *not* increase back to its initial value, but actually decreases further! The intrinsic pressure derivative for Tl-2201 is thus positive $(dT_c/dP)_{intr} > 0$. As seen in Fig. 14, if the sample is then annealed at progressively higher temperatures, each for 1 hour, $T_c$ relaxes back towards its initial value in a two-step fashion, indicating two distinct relaxation pathways. The low-temperature relaxation stagnates for temperatures near 110 K, but picks up again for temperatures above 180 K where a smaller high-temperature relaxation sets in. For Tl-2201, therefore, the measured pressure dependence of $T_c$ depends on the entire pressure/temperature history of the sample, $T_c = T_c(T, P, time)$. As one would expect, the importance of oxygen ordering effects in Tl-2201 depends strongly on the oxygen defect concentration [144].

Pressure-induced oxygen ordering effects have been observed on numerous other HTSC, including Hg-1201, Nd-123, Gd-123, $TlSr_2CaCu_2O_{7-y}$, $Sr_2CuO_2F_{2+y}$, and superoxygenated $La_2CuO_{4+y}$ [148, 149]. The activation energies for oxygen diffusion in Tl-2201, Y-123, and Hg-1201 were found by Sadewasser *et al.* [128] to increase with pressure, as expected; the activation volumes obtained allow an estimate of the most probable diffusion pathways for oxygen defects through the respective HTSC lattice. For further discussion of oxygen ordering effects in $La_{2-x}Sr_xCuO_4$ and other HTSC see the recent review by Lorenz and Chu [75].

From the above discussion it is apparent that oxygen ordering effects must be suppressed before the intrinsic pressure dependence $T_c^{intr}(P)$ can be established. There are three known ways to accomplish this: (1) carry out the entire experiment at sufficiently low temperatures that oxygen ordering effects are frozen out; (2) determine the initial pressure dependence $dT_c/dP$ only on optimally doped samples since at the extremum $T_c(n = n_{opt}) = T_c^{\max}$ the additional pressure-induced charge transfer from the oxygen ordering will have no effect; (3) study samples either with no mobile oxygen defects or with the maximum number of oxygen defects so that no empty defect sites are left.

The method (3) above was employed in the beautiful specific heat experiments to 10 GPa by Lortz *et al.* [134] on a fully oxygenated overdoped $YBa_2Cu_3O_7$ sample; the measurement of such a basic thermodynamic property as the specific heat allows the determination of the pressure dependence not only of the transition temperature $T_c(P)$ but also of the superconducting condensation energy $U_o(P)$, as seen in Fig. 15. For comparable change in $T_c$, the observed change in $U_o$ for underdoped YBCO is three times larger, reflecting the presence of superconducting fluctuation or pseudogap effects. In addition, from these results the pressure derivative of the carrier concentration is estimated to be $dn/dP \approx +0.0018$ to $+0.0026$ holes $Cu^{-1}GPa^{-1}$.

Sadewasser *et al.* [128] applied method (1) above to suppress oxygen ordering effects in an extensive study of Y-123 at different doping levels by maintaining the sample at temperatures below 200 K during the entire experiment in a He-loaded diamond-anvil cell. The results are shown in Fig. 11(left). Disappointingly, no simple systematics in $T_c^{intr}(P)$ are evident in these data. The "Modified Charge Transfer Model" as outlined above is unable to account for the data. Y-123 is evidently a VERY complex system, even without oxygen ordering effects. The presence of variably doped chains in Y-123 evidently adds a considerable (and unnecessary!) level of complexity. Y-123 and Y-124 are the only HTSC with CuO chains. To make advances in our understanding of the origins of HTSC, it is essential to study in depth the simplest systems possible. The tetragonal Hg-compounds, which exhibit relatively weak oxygen ordering effects, appear to be particularly attractive for further detailed studies and comparison with theory.

Very recently Tomita *et al.* [150, 151, 152] have carried out extensive studies of the critical current density $J_c$ across single grain boundaries in bicrystalline Y-123 samples for various oxygen concentrations and grain boundary mismatch angles. In all cases $J_c$ increases markedly with pressure. Interestingly, $J_c$ also exhibits relaxation effects following pressures changes at ambient temperature; the relaxation time is shorter than that for $T_c$, consistent with the usual picture that oxygen defects have a higher mobility in the grain boundary than in the bulk. That $J_c$ exhibits relaxation effects at all is evidence that some oxygen defect sites in the grain boundary must be vacant, i.e. high pressure experiments can be used as a probe to test whether the grain boundaries are fully oxygenated ot not [151]. Since it has been shown for Y-123 that $J_c$ increases with oxygen content [152], irrespective of the doping level, further enhancements in $J_c$ should be possible if all vacant sites can be filled with oxygen, for example, by subjecting the sample to a pure oxygen atmosphere at elevated temperatures and pressures.

#### 4.0.4 Intrinsic Pressure Dependence $T_c^{intr}(P)$

In spite of the great complexity of HTSC materials, a number of empirical guidelines have been identified [153] for enhancing the value of $T_c$: (1) vary the carrier concentration $n$ in the $CuO_2$ planes until its optimal value is reached (see Fig. 10); (2) increase the number of $CuO_2$ planes which lie close together (in a packet) in the oxide structure while maintaining optimal doping - "healthy" one-plane systems, like Tl-2201, have $T_c$ values in the range 90 - 100 K, two-plane systems in the range 100 - 120 K, and three-plane systems in the range 120 - 140 K; (3) try to position defects as far from the $CuO_2$ planes as possible; and (4) since $T_c$ is diminished with increasing buckling angle in the $CuO_2$ planes, develop structures where the $CuO_2$ planes are as flat as possible. Note that according to the above, the system Y-123 with $T_c^{\max} \simeq 92$ K is not particularly "healthy".

We now pose the question: what can we learn from high pressure experiments about how to further enhance the value of $T_c$? To answer this question, we should carefully select the systems we choose for experimentation, preferably picking "healthy" HTSC systems with relatively high values of $T_c$. Experimentation on "pathological" low-$T_c$ systems results in numerous factors changing at the same time, making the interpretation difficult if not impossible. The single-layer La-214 oxides are examples of such "pathological" systems, only possessing $T_c$ values in the range 30 - 40 K, far below the 90 - 100 K expected according to the above criteria for "healthy" single-layer systems such as Hg-1201 and Tl-2201. It is thus not surprising that in the La-214 oxides $T_c$ increases relatively rapidly with pressure as the structural distortions, which result in considerable buckling in the $CuO_2$ planes, are diminished. The La-214 systems are thus not suitable for further studies aimed at determining $T_c^{intr}(P)$. Similar structural transition effects led to early reports that the rate of increase of the transition temperature in HTSC with pressure, $|d\ln T_c/dP|$, is inversely related to the value of $T_c$ [154, 155, 120]. A closer examination of the relevant data to exclude systems with structural transitions, however, gave no evidence for such a correlation [17]. For further discussion we will focus on HTSC systems, like the one-, two- or three-layer Tl- or Hg-oxides or the two- or three-layer Bi-oxides, which are free of structural transition issues.

From the measured pressure dependences $T_c(P)$ for these systems, we would like to extract $T_c^{intr}(P)$, the "intrinsic" pressure dependence of $T_c$ for a given fixed carrier concentration $n$. This separation is extremely difficult for arbitrary doping levels since $n$ generally increases under pressure and $T_c$ is a particularly sensitive function of $n$, as seen in Fig. 10. Such a separation has been attempted for Hg-1201 [156] and a Tl-1212 compound [119] under strong simplying assumptions; such studies will only become really quantitative if the pressure dependence of $n$ is determined independently over the entire range of doping and pressure. Fortunately, for one value of $n$, namely, $n = n_{opt}$, the separation becomes simple, at least for the initial slope $dT_c/dP$, since at this extremum of $T_c(n)$ we have in Eq (10) $dT_c/dn = 0$ so that we obtain simply $dT_c/dP = (dT_c/dP)_{intr}$. Restricting our attention to optimally doped samples has the great advantage that the initial slope $dT_c/dP$ is free from the influence of changes in the carrier concentration $n$, and, as a bonus, oxygen ordering effects play no important role since they affect $T_c$ primarily through their influence on $n$.

For these reasons we now focus our attention on "healthy" optimally doped HTSC systems. The Hg-compounds are of particular interest here since their superconducting and structural properties have been studied on the same samples to high accuracy under purely hydrostatic pressure conditions in dense helium [8], as well as under quasihydrostatic pressures above 40 GPa (see Fig. 8) [9]. The optimally doped one-, two-, and three-layer Hg-compounds Hg-1201, Hg-1212, and Hg-1223 have, respectively, $T_c(0)$ values of 94 K, 127 K, and 134 K, initial pressure derivatives $dT_c/dP = +1.75 \pm 0.05$ K GPa$^{-1}$ for all three, relative pressure derivatives $d\ln T_c/dP \simeq +17.6$, +14.2, and $+12.9{\times}10^{-3}$ GPa$^{-1}$, and bulk moduli $B \simeq 69.4$, 84.0, and 92.6 GPa to 1.4% accuracy [157]. From these values the relative volume derivatives can be accurately determined to be $d\ln T_c/d\ln V = -B(d\ln T_c/dP) \simeq -1.22 \pm 0.05$, $-1.19 \pm 0.06$, and $-1.20 \pm 0.05$. It is quite remarkable that the relative pressure derivatives $d\ln T_c/dP$ differ by more than 30%, whereas the relative volume derivatives $d\ln T_c/d\ln V$, which would be expected to be of more direct physical relevance [17], turn out to be *identical* for all three Hg-compounds! This invariancy of the relative volume derivative gives strong evidence that the superconducting state, including the pairing mechanism, in the one-, two-, and three-layer Hg-compounds is the same. If one understands the nature of the superconductivity, and the mechanism(s) responsible for it, in the one-layer compound Hg-1201, one understands these basic properties in all three. This conclusion is underscored by the fact that the pressure dependence $T_c(P)$ to 40 GPa is nearly the same for all three Hg-compounds [9], as seen in Fig. 8.

We now consider the values of the relative volume derivative $d\ln T_c/d\ln V$ for further optimally doped HTSC systems: Y-123 ($-1.25 \pm 0.06$), Tl-2201 ($-1.35 \pm 0.4$), Tl-2212 ($-0.9 \pm 0.2$), Tl-2223 ($-1.16 \pm 0.3$), Bi-2212 ($-1.04 \pm 0.15$), and Bi-2223 ($-1.36$) [8, 146, 158]; the bulk modulus is known to lesser accuracy for the Tl- and Bi-systems than for Y-123 and the Hg-compounds. It is indeed remarkable that for all these optimally doped HTSC systems the intrinsic relative volume derivative turns out to be nearly the same $d \ln T_c / d \ln V \approx -1.2$, corresponding to the volume dependence

$$T_c \propto V^{-1.2}. \tag{14}$$

This is strong evidence that the nature of the superconductivity, and the mechanism(s) responsible for it, are the same in all high-$T_c$ cuprate superconductors. We note that this HTSC volume derivative $d \ln T_c / d \ln V \approx -1.2$ has the opposite sign, and is much weaker in magnitude, than the volume derivatives $d \ln T_c / d \ln V \simeq +7.2$ for Sn and +4.16 for $MgB_2$ which we discussed above. This fact by itself does not imply, however, that the electron-phonon interaction plays no role in HTSC. Negative volume derivatives are found in a number of transition metal systems, like La, Y, Lu, Sc or V (see Fig. 5) where the superconductivity is believed to be phonon mediated.

We are now in a position to understand why in the optimally doped Hg-compounds Hg-1201, Hg-1212, and Hg-1223 $T_c$ increases with pressure over such a relatively wide pressure range, resulting in the highest values of $T_c$ at 30 GPa for any known superconductor with the same number of $CuO_2$ layers. The very weak increase in the carrier concentration $n$ under pressure measured for Hg-1201 [159], and calculated for Hg-1223 [160], means that a relatively high pressure is required to increase $n$ sufficiently in the $T_c(n)$ phase diagram in Fig. 10 that the negative slope $(dT_c/dn)(dn/dP)$ becomes equal to the intrinsic positive slope $(dT_c/dP)_{intr} \simeq +1.75$ K GPa$^{-1}$, at which point $T_c(P)$ passes through a maximum at $P \approx 30$ GPa. This maximum value $T_c^{\max}$(30 GPa), can be estimated from Eq (11) for Hg-1223 by setting $T_c^{\max}(0) = 134$ K and $dT_c^{\max}/dP \simeq +1.75$ K GPa$^{-1}$ and assuming $n_{opt}$ and $T_c^{\max}$ are independent of $P$ and $n$, respectively. If one now asks what value of $dn/dP$ is required that $T_c(P)$ reaches its maximum value, where $dT_c/dP = 0$, at 30 GPa, out comes $dn/dP \simeq +0.00129$ hole GPa$^{-1}$. If this value of $dn/dP$ is then inserted in Eq (11), then one obtains the estimate $T_c^{\max}$(30 GPa) $\simeq 163$ K which is close to the measured value (see Fig. 8). This value of $dn/dP$, which is somewhat smaller than that estimated for Y-123, agrees reasonably well with a calculation by Singh *et al.* [160]. Thermopower measurements by Chen *et al.* [159] indicate an even smaller value. If, as suggested by Xiong *et al.* [161], $\beta = 50$ is substituted for $\beta = 82.6$ in the Tallon formula, $dn/dP \simeq +0.00166$ hole/GPa is obtained, but the estimate $T_c$(30 GPa) $\simeq 163$ K remains the same.

We would now like to explore the question as to the origin of the relatively weak dependence of $T_c$ on sample volume $T_c \propto V^{-1.2}$ in HTSC materials. When hydrostatic pressure is applied to a HTSC, the unit cell is compressed in all three directions. However, with the exception of the La-214 compound family, the compressibility in the direction perpendicular to the $CuO_2$ planes, the $c$-direction, is in general approximately twice as large as in a direction parallel to the $CuO_2$ planes [17]. The central question at hand is: does the intrinsic increase of $T_c$ with pressure, reflected in $T_c \propto V^{-1.2}$, originate primarily from the reduction in the *separation* between the $CuO_2$ planes or from the reduction in the *area* of these planes? To answer this question, we must turn to uniaxial pressure experiments which have the potential to unravel the information hidden in the hydrostatic pressure studies.

### 4.0.5 Uniaxial Pressure Results

Uniaxial pressure experiments are technologically very difficult and require high quality single crystals of sufficient size. The partial pressure derivatives along the crystallographic axes $dT_c/dP_a$, $dT_c/dP_b$, and $dT_c/dP_c$ can be determined either by applying force directly to the crystal along the respective crystallographic directions [162], or through combined ultrahigh resolution thermal expansion and specific heat measurements using the Ehrenfest relation $dT_c/dP_i = \Delta\alpha_i V_m T_c/\Delta C_p$, where $\Delta\alpha_i$ and $\Delta C_p$ are the mean-field jumps of the thermal expansion coefficient and specific heat, respectively, and $V_m$ is the molar volume [163]. Note that the hydrostatic pressure derivative can be written as the sum of the respective partial pressure derivatives $dT_c/dP = dT_c/dP_a + dT_c/dP_b + dT_c/dP_c$. The result of the "Modified Charge Transfer Model" in Eq (11) can be applied by simply replacing $dT_c/dP$ by the respective partial pressure derivative $dT_c/dP_i$ where $i = a, b, c$.

The results of detailed thermal expansion studies by Meingast *et al.* [127] on crystals from the $Y_{1-y}Ca_yBa_2Cu_3O_x$ compound series are shown in Fig. 16. At ambient pressure $T_c(n)$ is seen to pass through a maximum at $T_c \simeq 93$ K for $n = n_{opt} \simeq 0.16$ K. The partial pressure derivatives generally change from positive to negative as the carrier concentration $n$ increases, reflecting the influence of pressure-induced charge transfer. At optimal doping one has $n = n_{opt}$ and $dT_c/dn = 0$ so that the partial pressure derivatives give the intrinsic effect directly. In Fig. 16(c) we see that at optimal doping $dT_c/dP_c \approx 0$; this implies that enhancing the interplanar coupling by pushing the $CuO_2$ planes closer together has no measureable effect on the superconducting state. This, together with the fact that $dT_c/dP_c$ depends linearly on $n$, in agreement with Eq (13), gives strong evidence that the primary effect of compression in the $c$ direction is to enhance the the carrier concentration $n$.

On the other hand, as seen in Fig. 16(b), compressing the $CuO_2$ planes themselves enhances $T_c$ at the rate $\sim +1$ K GPa$^{-1}$, in good agreement with hydrostatic pressure studies on the same compound series [126]. Parallel thermal expansion studies [163] on an optimally doped detwinned $YBa_2Cu_3O_x$ crystal give the following partial pressure derivatives: $dT_c/dP_a \simeq$ -1.9 K GPa$^{-1}$, $dT_c/dP_b \simeq$ +2.2 K GPa$^{-1}$, and $dT_c/dP_c \simeq 0$ K GPa$^{-1}$, in excellent agreement with later studies by Kund and Andres [164] as well as with direct uniaxial pressure experiments by Welp *et al.* [162]. All these studies confirm that the intrinsic pressure effect within the $CuO_2$ planes is large, in contrast to the negligible effect along the $c$ axis perpendicular to these planes. The opposite sign of the partial pressure derivatives in the $a$ and $b$ directions is simply a reflection of the above $T_c-$optimization rule (4) whereby the $CuO_2$ planes should be as flat (therefore tetragonal) as possible to maximize $T_c$; Chen *et al.* [165] have developed a model which accounts for the $dT_c/dP_i$ anisotropies in terms of anisotropies in both the hole dispersion and the pairing interaction.

In the above experiments, a compression along one axis (unfortunately) leads to an expansion along the other two axes, so that all three change. The partial pressure derivatives, however, can be converted into the partial strain derivatives $dT_c/d\epsilon_a$, $dT_c/d\epsilon_b$, and $dT_c/d\epsilon_c$, if the elastic constants are known to sufficient accuracy. For Y-123 the dominant strain derivative at optimal doping turns out to be $dT_c/d\epsilon_b$, the other two being at least 5× smaller [166]. When pressure is applied to a Y-123 crystal, the intrinsic effect on $T_c$ is predominantly caused by a strain in the $CuO_2$ plane along the $b$ (chain) direction. On the other hand, in the double-chain system Y-124 compression along the a-direction is dominant [167]. Studies of bond-length systematics in $RBa_2Cu_4O_8$ across a portion of the rare-earth series R both at ambient [168] and high pressure [169] have revealed that $T_c$ correlates well with the Cu-O bond lengths within, rather than perpendicular to, the $CuO_2$ planes.

The above results underscore the considerable complexity of the two HTSC compounds, Y-123 and Y-124, which contain CuO chains, a superfluous structural element unnecessary for high-$T_c$ superconductivity. Indeed, the HTSC systems with the highest values of $T_c$ have no chains. To make significant progress in our understanding of the basic issues regarding superconductivity in HTSC, one would be well advised to focus on tetragonal systems free from CuO chain structures.

High-resolution thermal expansion experiments have been carried out on a nearly optimally doped Bi-2212 crystal by Meingast *et al.* [170] with results: $dT_c/dP_i \simeq$ +1.6, +2.0, and -2.8 K $GPa^{-1}$ for $i = a, b, c$. Kierspel *et al.* [171] obtain somewhat different values for Bi-2212: $dT_c/dP_i \simeq +0.9$, +0.9, and < 0.4 K $GPa^{-1}$, respectively, yielding a total pressure derivative $dT_c/dP \approx 2$ K $GPa^{-1}$, in good agreement with the hydrostatic pressure dependence [172].

Unfortunately, uniaxial pressure results have yet to be published for the tetragonal Hg- and Tl-compound families due to the difficulty in obtaining high quality crystals of sufficient size. Further experimentation on the Hg cuprates in particular is strongly encouraged since these oxides are blessed with a relatively simple structure and thus offer an excellent opportunity for obtaining definitive results.

# 5 Conclusions and Outlook

Taken together, the above experiments support the picture that the dimensions of the $CuO_2$ planes, rather than the separation between them, primarily determines the maximum value of $T_c$ in a given HTSC: the closer the planes are to being square and flat, and the smaller their area, the higher the value of $T_c$. A similar conclusion regarding the relative importance of the the in-plane and out-of-plane lattice parameters was reached in a review by Schilling and Klotz [17] in 1991 and in a paper by Wijngaarden *et al.* [119] in 1999 who commented that "There is quite some evidence that $c$ mainly influences doping, while $a$ mainly influences the intrinsic $T_c$". The high-pressure experiments carried out to-date thus lend support to those theories where the interactions within the $CuO_2$ planes are primarily responsible for the superconducting pairing.

From this it follows that the ubiquitous intrinsic volume dependence $T_c \propto V^{-1.2}$ for nearly optimally doped HTSC arises from the compression of the $CuO_2$ planes, rather than from a reduction in the separation between them. To obtain the intrinsic dependence of $T_c$ on the in-plane lattice parameter $a$, we evaluate $d\ln T_c/d\ln a = -\kappa_a^{-1}(d\ln T_c/dP)$, where $\kappa_a \equiv -d\ln a/dP$ is the $a$-axis compressibility. Using the above values of $d\ln T_c/dP$ for the one-, two-, and three-layer Hg-compounds as well as the $\kappa_a$-values 4.26, 2.94, and $2.57\times10^{-3}$ $\text{GPa}^{-1}$ [173], respectively, we obtain $d\ln T_c/d\ln a$ = -4.1, -4.8, and -5.0 which translates into the approximate in-plane lattice parameter dependence

$$T_c \propto a^{-\delta}, \text{ where } \delta = 4.5 \pm 0.5. \tag{15}$$

This expression implies that at optimal doping the intrinsic $T_c$ is roughly proportional to the inverse square of the area $A$ of the $CuO_2$ planes, $T_c \propto A^{-2}$. Similar results are obtained for other optimally doped HTSC systems.

This is one of the single most significant results to be distilled from high pressure experiments on HTSC materials and is information not readily available through other means. Besides giving information on the superconducting mechanism, this dependence points to an additional strategy for further increasing $T_c$. To the above five "$T_c$ optimization rules", we can now add: "(6) seek out structures which apply maximal compression to the $CuO_2$ planes without causing them to buckle." According to the above relations, if we were to apply sufficient pressure to an optimally doped Hg-1223 sample to compress its in-plane dimension by about 20%, without adding defects or increasing the number of charge carriers, $T_c$ should increase from 134 K to 304 K and we would have the world's first room temperature superconductor!

HTSC systems with the same number of $CuO_2$ planes generally have different values of $T_c = T_c^{\max}$ at optimal doping. It is interesting to inquire whether this difference arises from a variation in the in-plane lattice parameter $a$, i.e. $T_c^{\max} \propto a^{-4.5}$. From Fig. 17 one can see that no such simple correlation exists. The single-plane material with the highest value of $T_c$, Hg-1201 with $T_c^{\max} \simeq 98$ K, has the *largest* value of $a$, and the compound in Fig. 17 with the lowest value of $T_c$, $La_{1.85}Sr_{0.15}CuO_4$ with $T_c^{\max} \simeq 36$ K, has the *smallest* value of $a$. Perhaps $La_{1.85}Sr_{0.15}CuO_4$ owes its anomalously low value of $T_c$ to an overcompression of its $CuO_2$ plane, resulting in strong structural distortions and plane buckling, effects which are known to degrade $T_c$. The $a$ values of the other systems listed differ by only 1.4% which corresponds to $\sim$ 5 GPa or a change in $T_c$ by only $7-8$ K. Raising the compression level from 1.4% to 20% is a worthy goal but constitutes a very difficult challenge for materials scientists.

An important point remaining is to identify what information the above dependence $T_c \propto a^{-4.5}$ gives on the nature of the superconducting state in HTSC. If we assume two electrons are bound in a Cooper pair by electron-phonon, electron-electron, electron-magnon, or other effective interactions, $V_{eff}$, a BCS-like expression is appropriate for weak interactions

$$T_c \simeq \langle\omega\rangle \exp\left[-1/\mathcal{V}_{eff}N(E_f)\right], \tag{16}$$

where $\langle\omega\rangle$ is the characteristic energy of the intermediary bosons. Since both $\mathcal{V}_{eff}$ and $N(E_f)$ are in the exponent in Eq (16), it is likely that their pressure dependence is responsible for that of $T_c$. Early high-pressure measurements of the spin susceptibility of $La_{1.85}Sr_{0.15}CuO_4$ [175] and Y-123 [176] and band-structure calculations for Hg-1223 [160] found the changes in $N(E_f)$ under pressure to be less than 0.2, 0.1, and 0.5 %/GPa, respectively. For $La_{1.85}Sr_{0.15}CuO_4$ this change in $N(E_f)$ is too small to account for the rapid increase of $T_c$ under pressure; to make a similar evaluation for Y-123 and Hg-1223, where $d\ln T_c/dP$ is much smaller, the accuracy of the $dN(E_f)/dP$ determination would have to be considerably enhanced.

The questions remains: why does $\mathcal{V}_{eff}N(E_f)$ increase with pressure at a rate such that $T_c \propto a^{-4.5}$ ? Unfortunately, this $T_c(a)$ dependence alone gives insufficient information to allow one to unequivocally identify the pairing interaction. The intrinsic pressure dependence $dT_c/dP \approx +1$ to 2 K $GPa^{-1}$ for "healthy" HTSC easily falls within the wide range of dependences observed for transition metal superconductors (see discussion above) where electron-phonon pairing is well established. From an analysis of the high-pressure results on HTSC, Neumeier [177] has come to a similar conclusion, namely, electron-phonon coupling is a possible pairing interaction for HTSC.

The above analysis leading to the intrinsic relation $T_c \propto a^{-4.5}$ has, unfortunately, only been carried out on HTSC near optimal doping. This restriction was necessitated by the need to eliminate pressure-induced changes in $n$. To establish $T_c^{intr}(P)$ over a wide range of doping, an independent determination of $dn/dP$ over this entire range must first be carried out.

To shed light on the pairing interaction through high pressure studies, it will be necessary to combine $T_c(P)$ determinations under hydrostatic and uniaxial pressure with simultaneous measurements (preferably on the same crystal) of other important superconducting- and normal-state properties such as the superconducting gap, the pseudo-gap in the underdoped region, superconducting condensation energy, magnetic susceptibility, electrical resistivity, Hall effect, thermoelectric power, etc. Aronson *et al.* [178] made an early attempt along these lines by carrying out high-pressure Raman scattering studies on antiferromagnetic $La_2CuO_4$ and found that the superexchange interaction $J$ increases approximately as $J \propto a^6$. Such studies, if expanded to other HTSC, have the potential to test the viability of spin-fluctuation theories; measurements of the pressure-dependent magnetic susceptibility at elevated temperatures would provide similar information on $J(P)$. Further studies on the La-214 system would seem ill advised since the rampant structural distortions and transitions in this system make a quantitative analysis extremely difficult. It would be of considerable interest to attempt such studies on crystals from the Hg- and Tl-compound families over the full range of doping. Special emphasis should be placed on uniaxial pressure experiments since they can provide the kind of detailed information needed to make real progress.

**Acknowledgments.** The author would like to acknowledge research support by the National Science Foundation under grant DMR-0404505 and thank J. Hamlin for technical assistance.

# 6 Bibliography

| | | | | | | | | | | | | | | | | | |
|---|---|---|---|---|---|---|---|---|---|---|---|---|---|---|---|---|---|
| **H** | | | | | | | | | | | | | | | | | **He** |
| **Li**<br>14<br>30 | **Be**<br>0.026 | | | | | | | | | | | **B**<br>11<br>250 | **C** | **N** | **O**<br>0.6<br>100 | **F** | **Ne** |
| **Na** | **Mg** | | | | | | | | | | | **Al**<br>1.14 | **Si**<br>8.2<br>15.2 | **P**<br>13<br>30 | **S**<br>17.3<br>190 | **Cl** | **Ar** |
| **K** | **Ca**<br>15<br>150 | **Sc**<br>0.35<br>21 | **Ti**<br>0.39 | **V**<br>5.38<br>16.5<br>120 | **Cr** | **Mn** | **Fe**<br>2.1<br>21 | **Co** | **Ni** | **Cu** | **Zn**<br>0.875 | **Ga**<br>1.091<br>7<br>1.4 | **Ge**<br>5.35<br>11.5 | **As**<br>2.4<br>32 | **Se**<br>8<br>150 | **Br**<br>1.4<br>100 | **Kr** |
| **Rb** | **Sr**<br>4<br>50 | **Y**<br>19.5<br>115 | **Zr**<br>0.546<br>11<br>30 | **Nb**<br>9.50<br>9.9<br>10 | **Mo**<br>0.92 | **Tc**<br>7.77 | **Ru**<br>0.51 | **Rh**<br>.00033 | **Pd** | **Ag** | **Cd**<br>0.56 | **In**<br>3.404 | **Sn**<br>3.722<br>5.3<br>11.3 | **Sb**<br>3.9<br>25 | **Te**<br>7.5<br>35 | **I**<br>1.2<br>25 | **Xe** |
| **Cs**<br>1.3<br>12 | **Ba**<br>5<br>18 | insert<br>**La-Lu** | **Hf**<br>0.12<br>8.6<br>62 | **Ta**<br>4.483<br>4.5<br>43 | **W**<br>0.012 | **Re**<br>1.4 | **Os**<br>0.655 | **Ir**<br>0.14 | **Pt** | **Au** | **Hg-α**<br>4.153 | **Tl**<br>2.39 | **Pb**<br>7.193 | **Bi**<br>8.5<br>9.1 | **Po** | **At** | **Rn** |
| **Fr** | **Ra** | insert<br>**Ac-Lr** | **Rf** | **Ha** | | | | | | | | | | | | | |

ambient pressure superconductor: $T_c$(K), $T_c^{max}$(K), P(GPa)

high pressure superconductor: $T_c^{max}$(K), P(GPa)

| | | | | | | | | | | | | | | |
|---|---|---|---|---|---|---|---|---|---|---|---|---|---|---|
| **La**-fcc<br>6.00<br>13<br>15 | **Ce**<br>1.7<br>5 | **Pr** | **Nd** | **Pm** | **Sm** | **Eu** | **Gd** | **Tb** | **Dy** | **Ho** | **Er** | **Tm** | **Yb** | **Lu**<br>2.5<br>22 |
| **Ac** | **Th**<br>1.368 | **Pa**<br>1.4 | **U**<br>0.8(β)<br>2.4(α)<br>1.2 | **Np** | **Pu** | **Am**<br>0.79<br>2.2<br>6 | **Cm** | **Bk** | **Cf** | **Es** | **Fm** | **Md** | **No** | **Lr** |

**Figure 1.** Periodic Table listing 29 elements superconducting at ambient pressure (yellow) and 23 elements which only superconduct under high pressure (green). For each element the upper position gives the value of $T_c$(K) at ambient pressure; middle position gives maximum value $T_c^{\max}$(K) in a high-pressure experiment at $P$(GPa) (lower position). In many elements multiple phase transitions occur under pressure. If $T_c$ decreases under pressure, only the ambient pressure value of $T_c$ is given. Sources for $T_c$ values at ambient pressure are given in Ref. [13]. Sources for $T_c$ values under high pressure are given in Ref. [14].

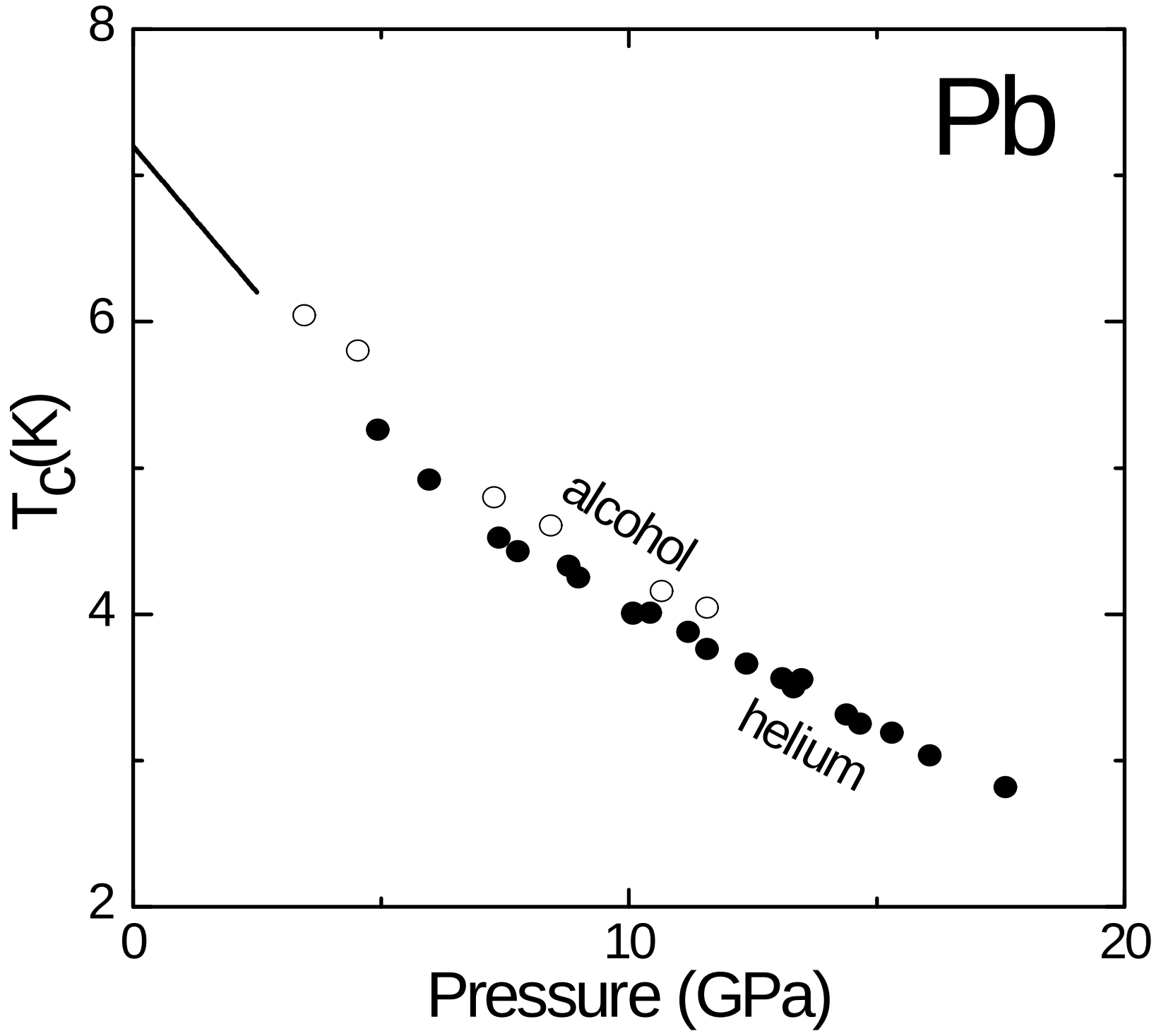

**Figure 2.** Pressure dependence of $T_c$ for Pb from Ref. [15] using helium (•) and methanol-ethanol (◦) pressure media. Straight line gives initial dependence from Ref. [18].

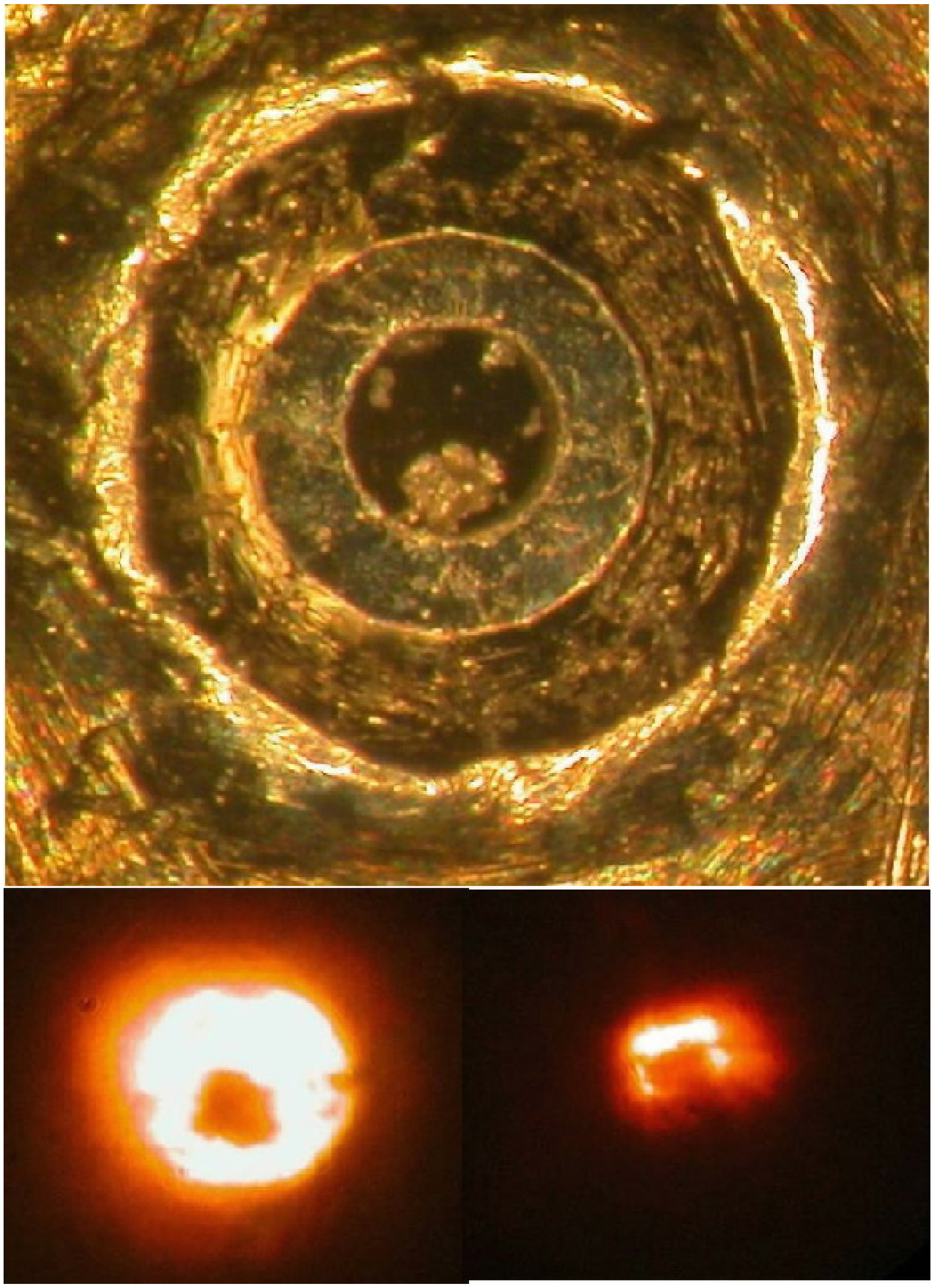

**Figure 3.** (top) Gold-plated rhenium gasket with 250 $\mu m$ dia hole containing Li sample. (bottom) Transmitted-light photograph of hole containing Li sample at (left) ambient pressure and (right) 30 GPa. Figure taken from Ref. [46].

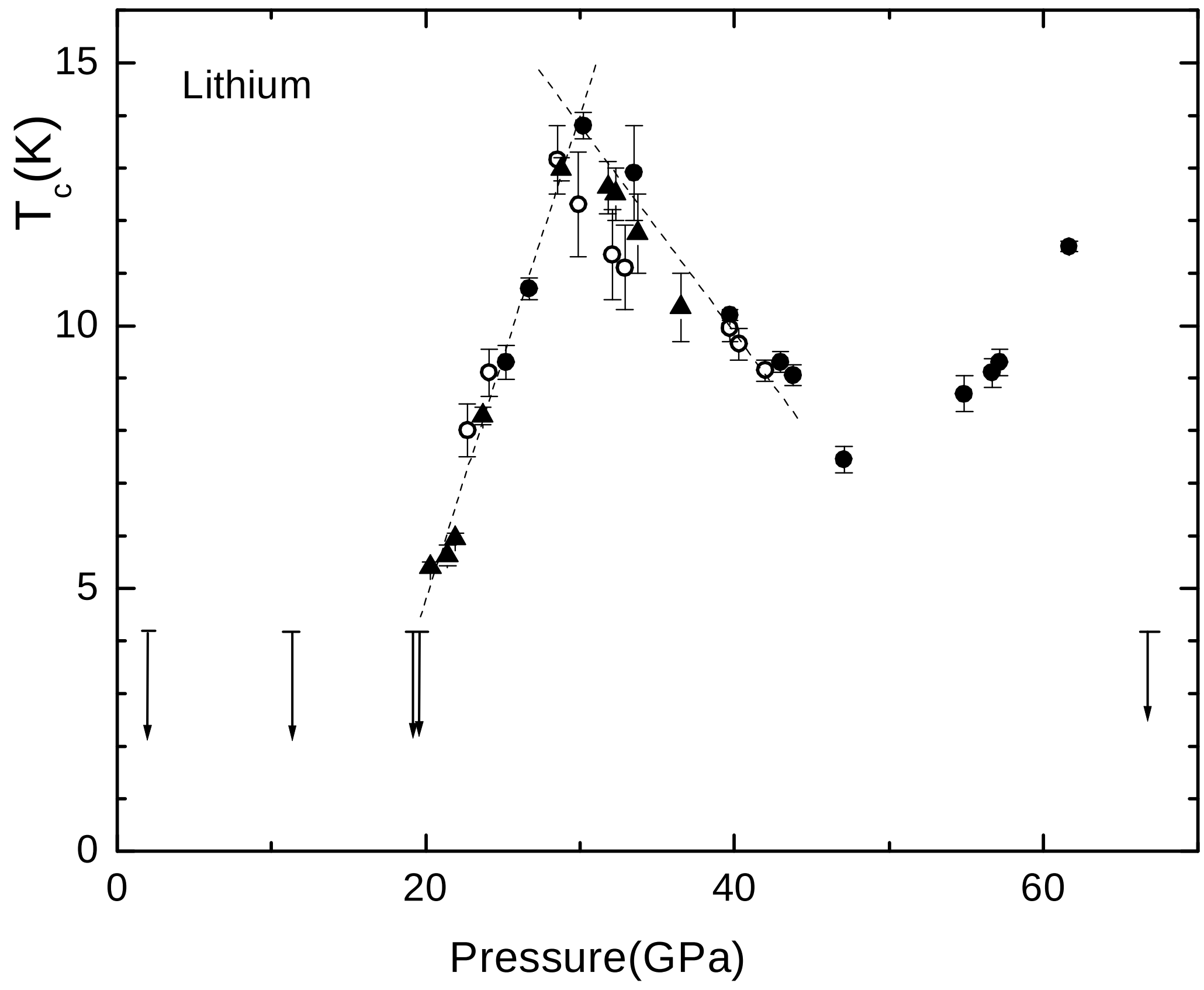

**Figure 4.** Superconducting phase diagram $T_c(P)$ of Li metal under nearly hydrostatic pressure (helium) to 67 GPa from Ref. [46]. Dashed lines are guides to eye. Several structural phase transitions are indicated.

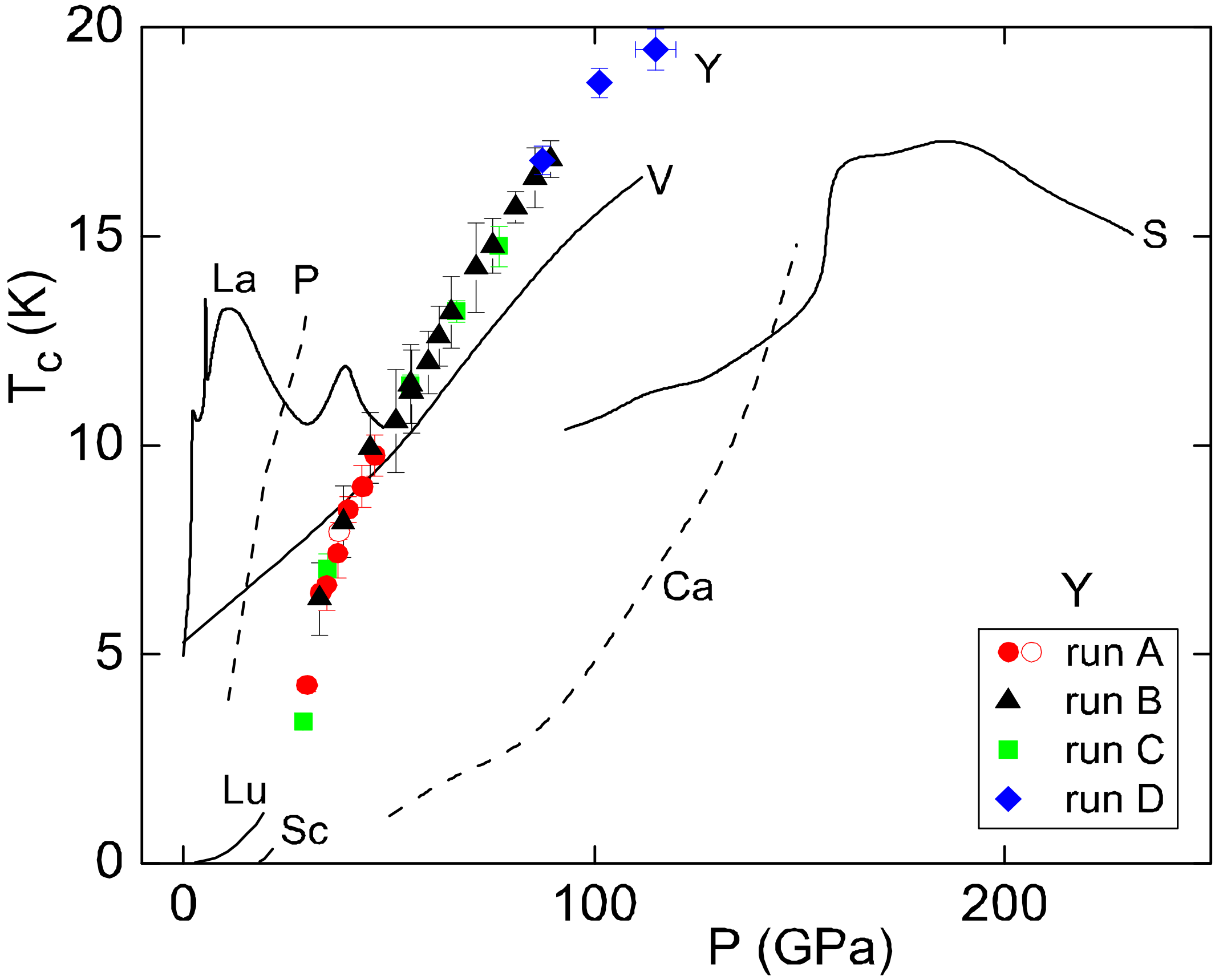

**Figure 5.** Superconducting phase diagrams $T_c(P)$ for non-alkali elements with the highest values of $T_c$ under pressure (Ca [65], La [66], P [67], S [68], V [69], Y [70, 71]) as well as for Lu [72] and Sc [73]. For clarity only the three highest-pressure data are shown for Y in run D.

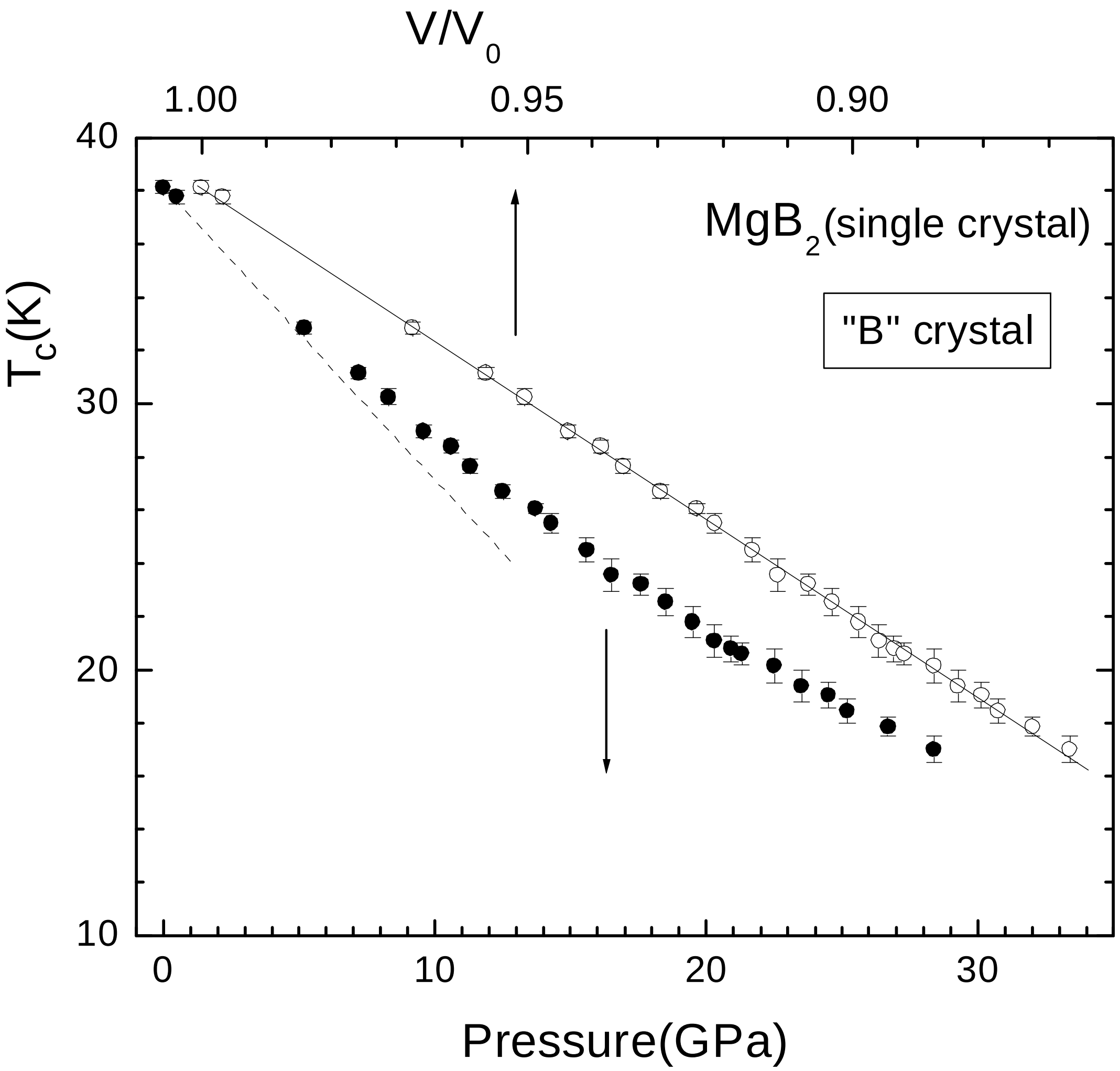

**Figure 6.** Dependence of $T_c$ for a $MgB_2$ single crystal on nearly hydrostatic pressure $P$ and relative volume $V/V_o$ in a He-loaded diamond-anvil cell, from Ref. [93].

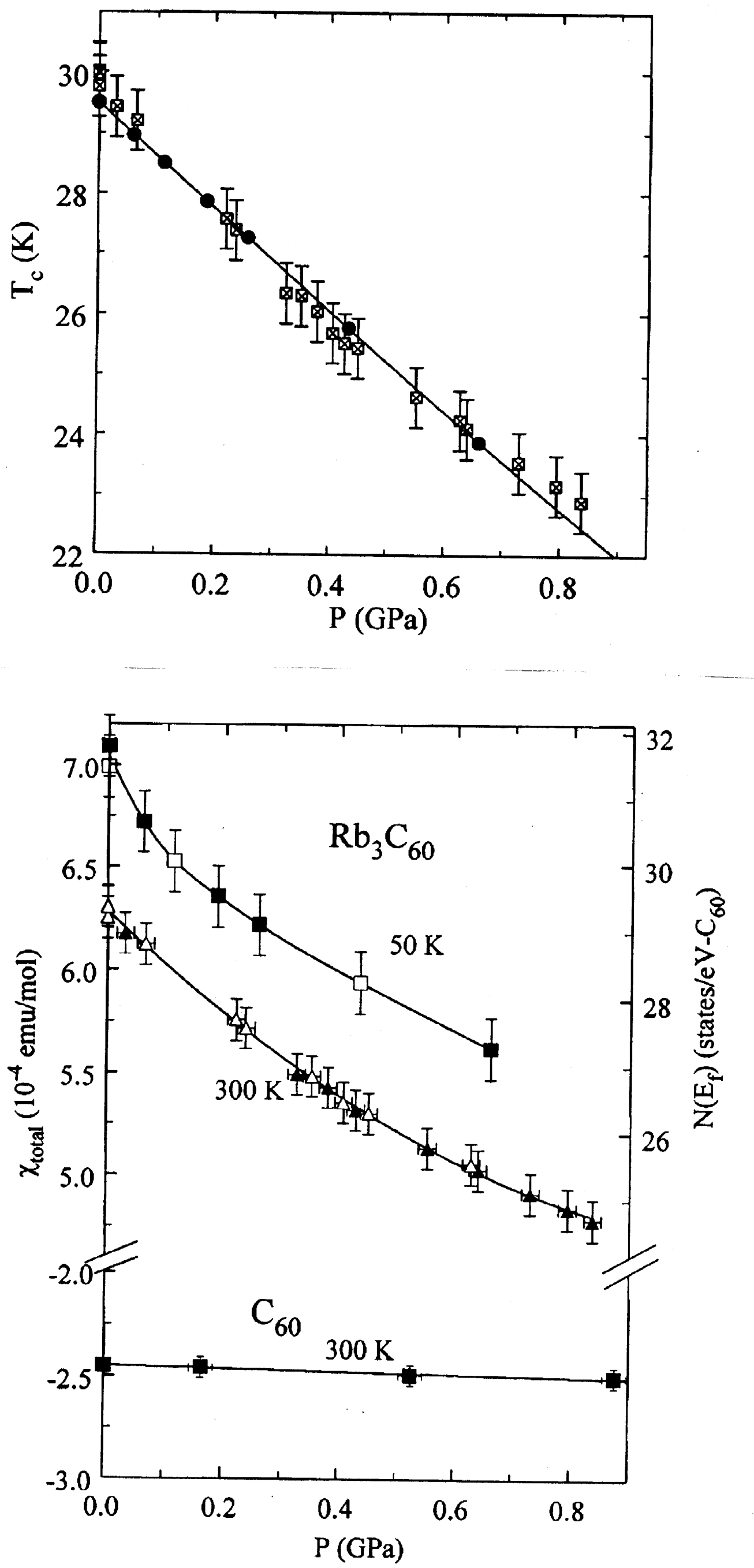

**Figure 7.** Results of hydrostatic pressure studies on $Rb_3C_{60}$ from Ref. [110]: (top) superconducting transition temperature $T_c$; (bottom) magnetic susceptibility and estimated electronic density of states $N(E_f)$ at 50 K and 300 K. Data for $C_{60}$ at 300 K are also shown. Both $T_c$ and $N(E_f)$ decrease rapidly with pressure.

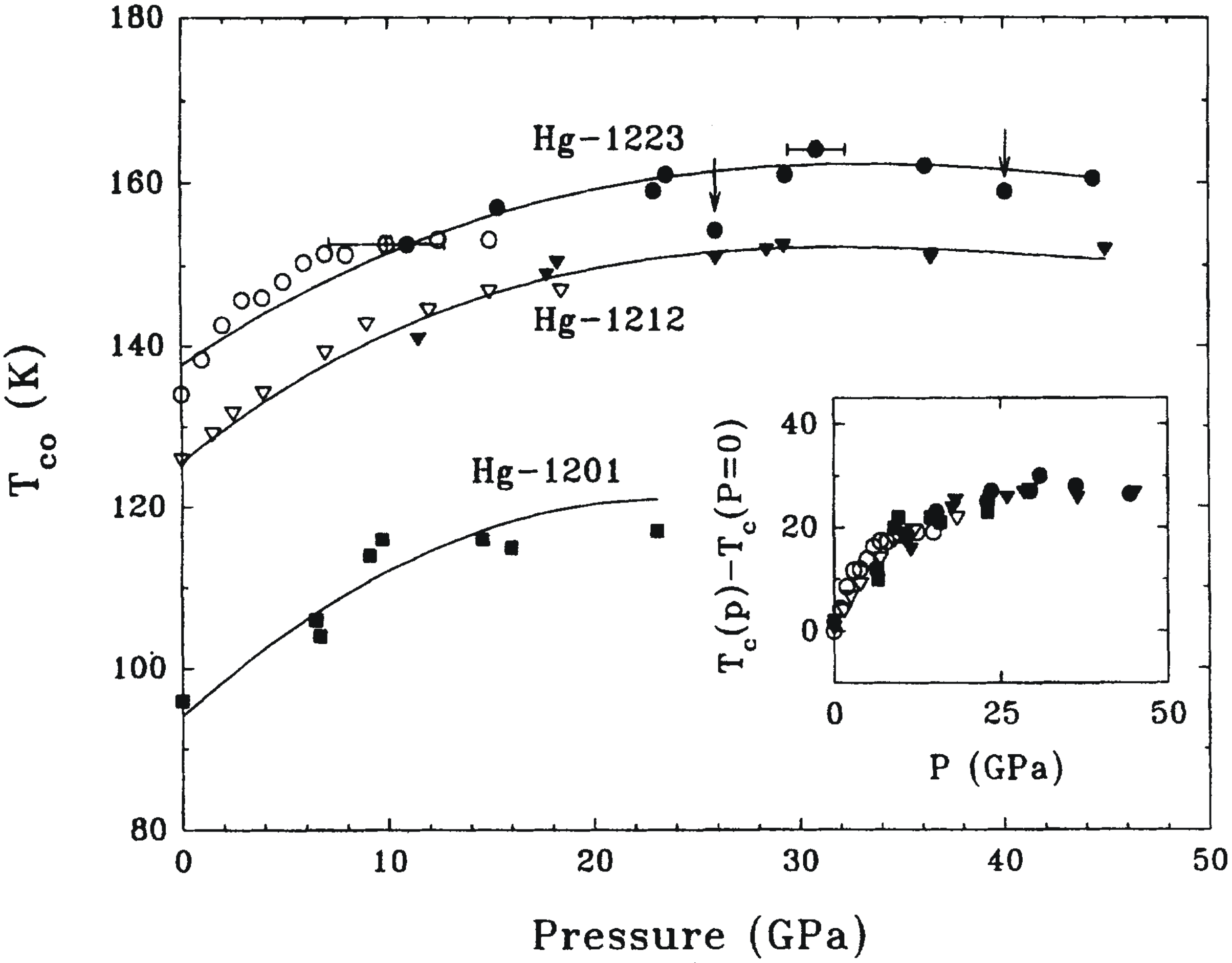

**Figure 8(left).** Hg-compounds: $T_c$ versus pressure to 45 GPa from Ref. [9].

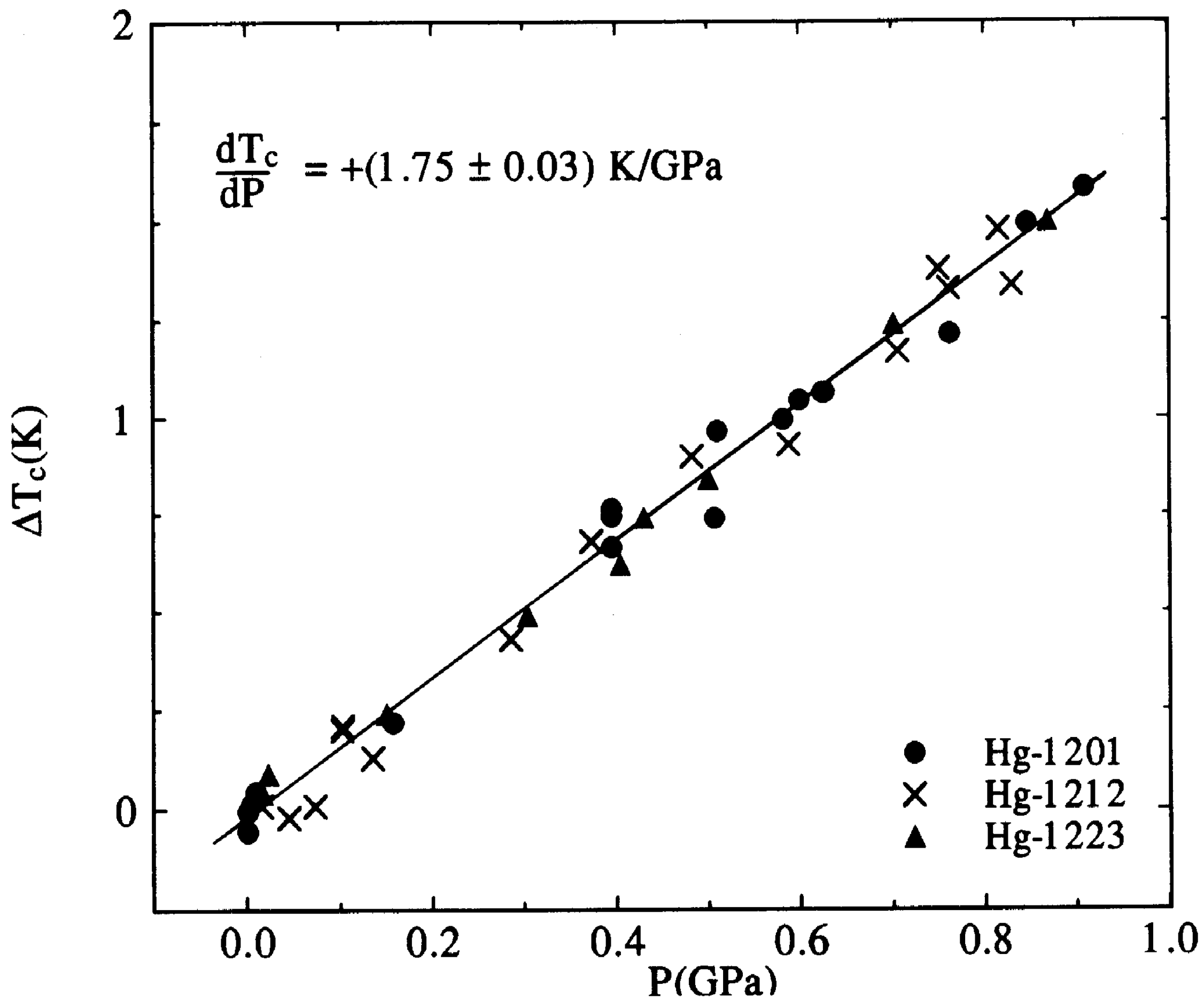

**Figure 8(right).** Hg-compounds: change in $T_c$ versus pressure to 0.9 GPa from Ref. [8]. The initial pressure dependence $dT_c/dP$ for all three Hg-compounds is identical.

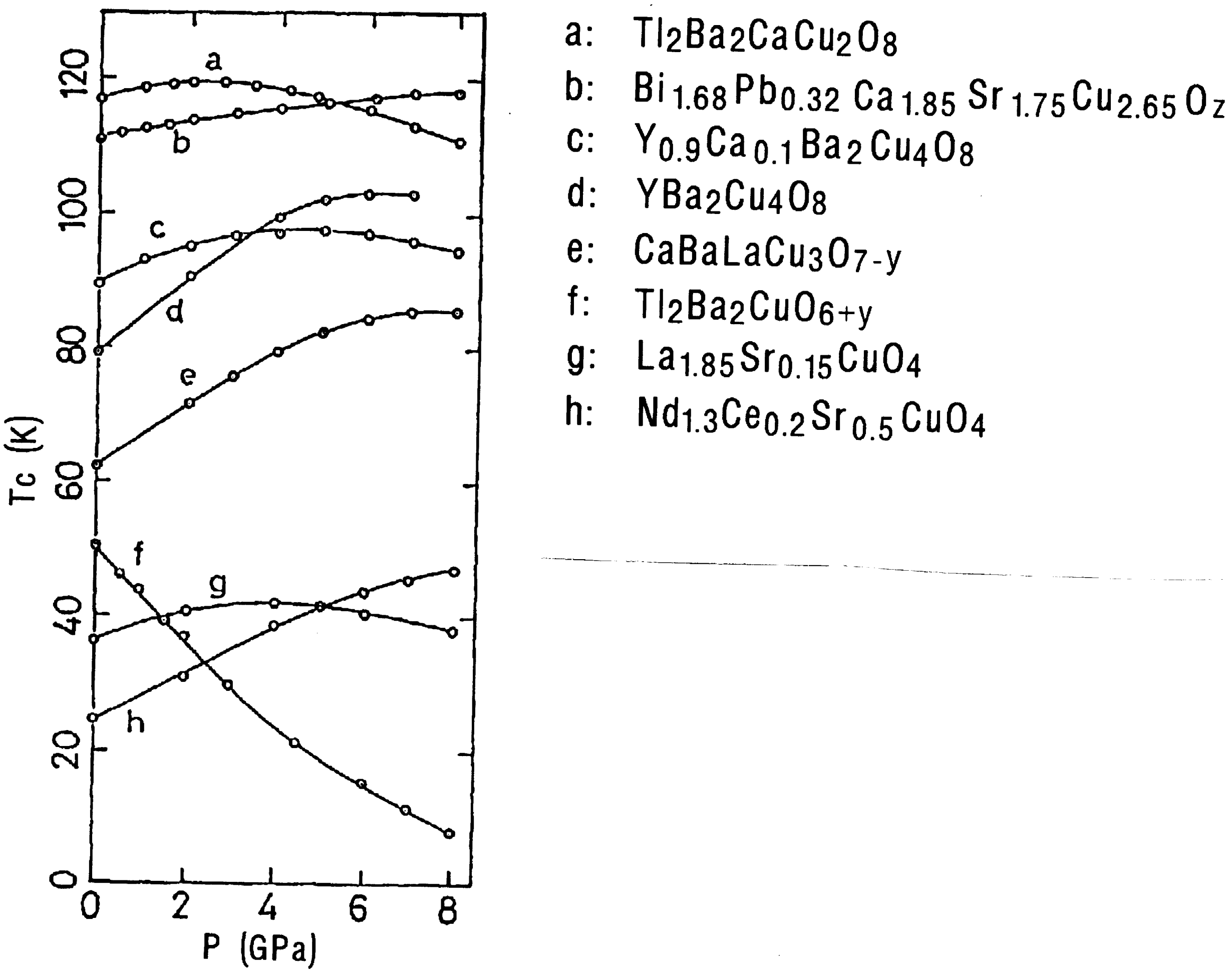

**Figure 9.** $T_c$ versus pressure for several HTSC. Figure taken from Ref. [122].

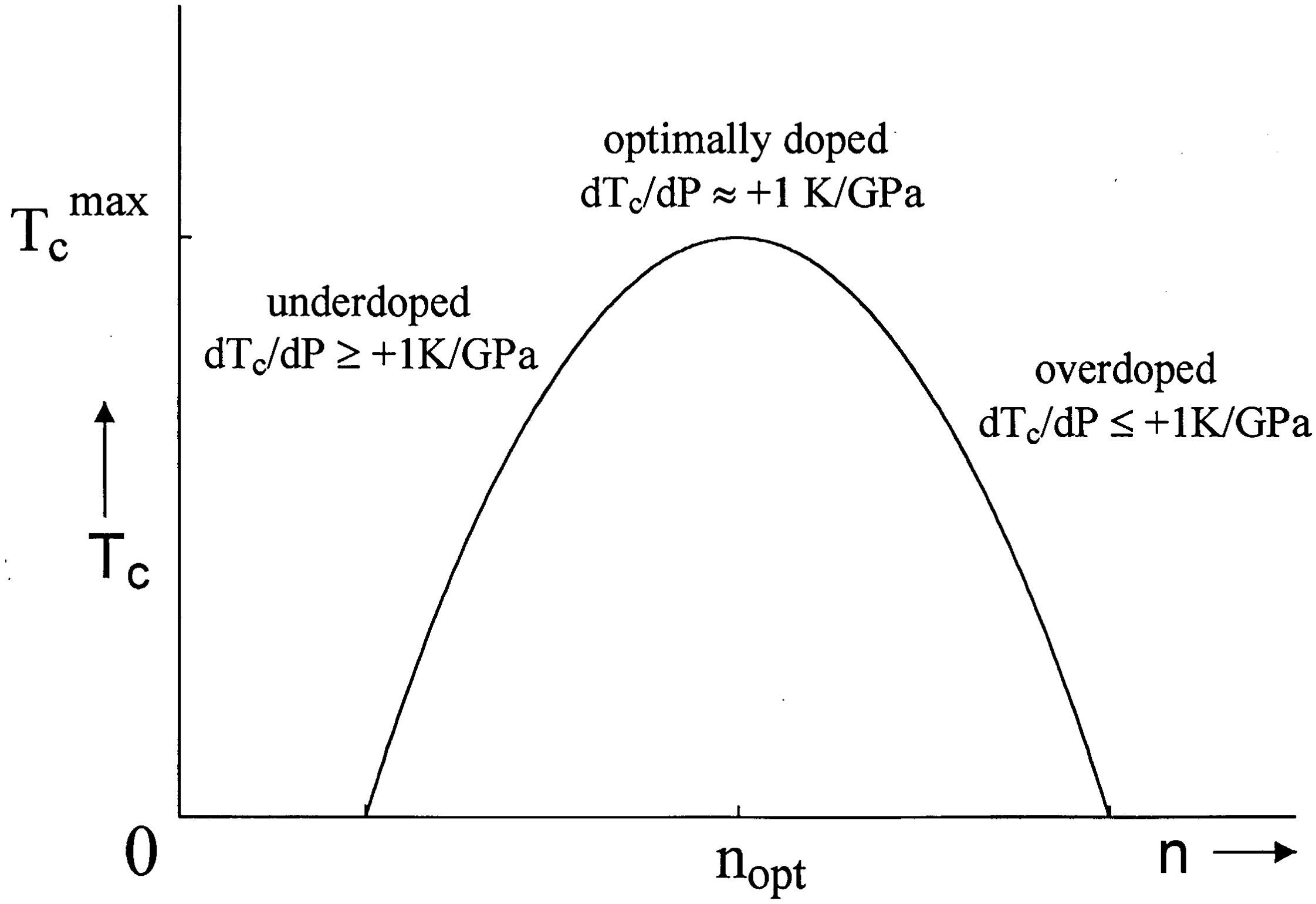

**Figure 10.** Canonical dependence of $T_c$ on carrier concentration $n$ for HTSC according to Eq 8. Typical values for $dT_c/dP$ in the underdoped, optimally doped, and overdoped regions are given.

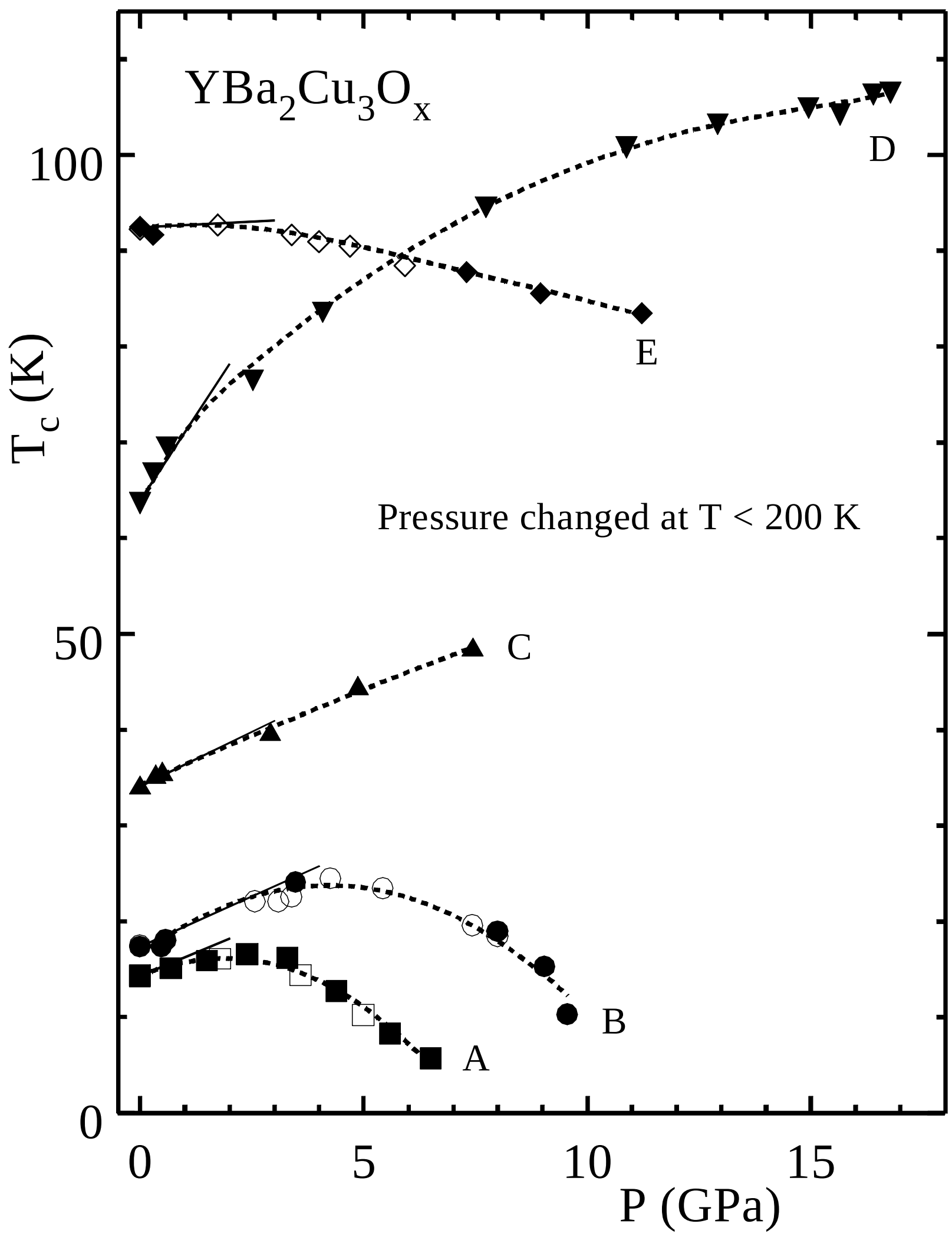

**Figure 11(left).** Dependence of $T_c$ on pressure from Ref. [128] for five $YBa_2Cu_3O_x$ samples $A \rightarrow E$ with increasing oxygen content $x$. See text for details.

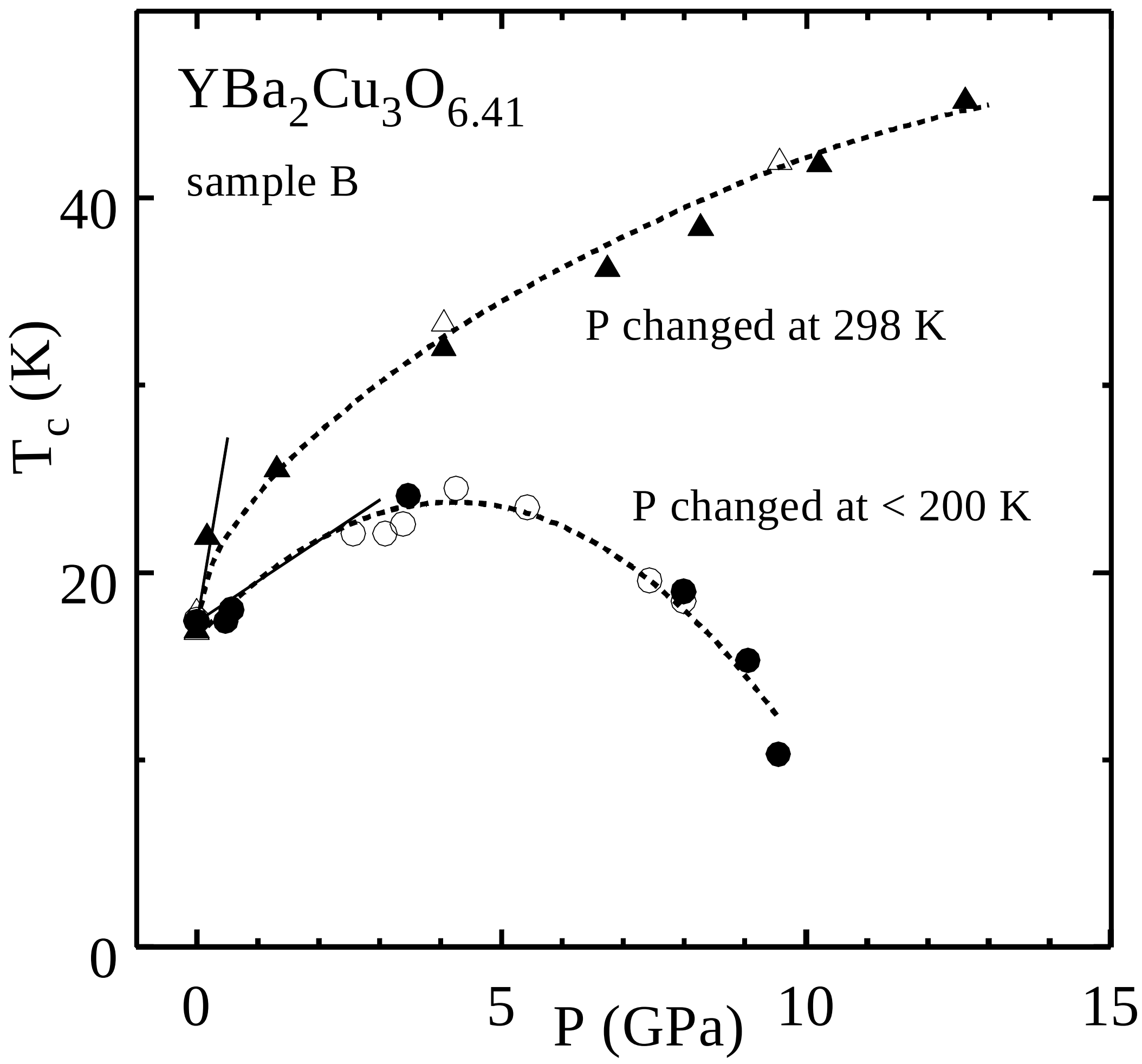

**Figure 11(right).** Dependence of $T_c$ on pressure from Ref. [128] for the underdoped sample $B$. See text for details.

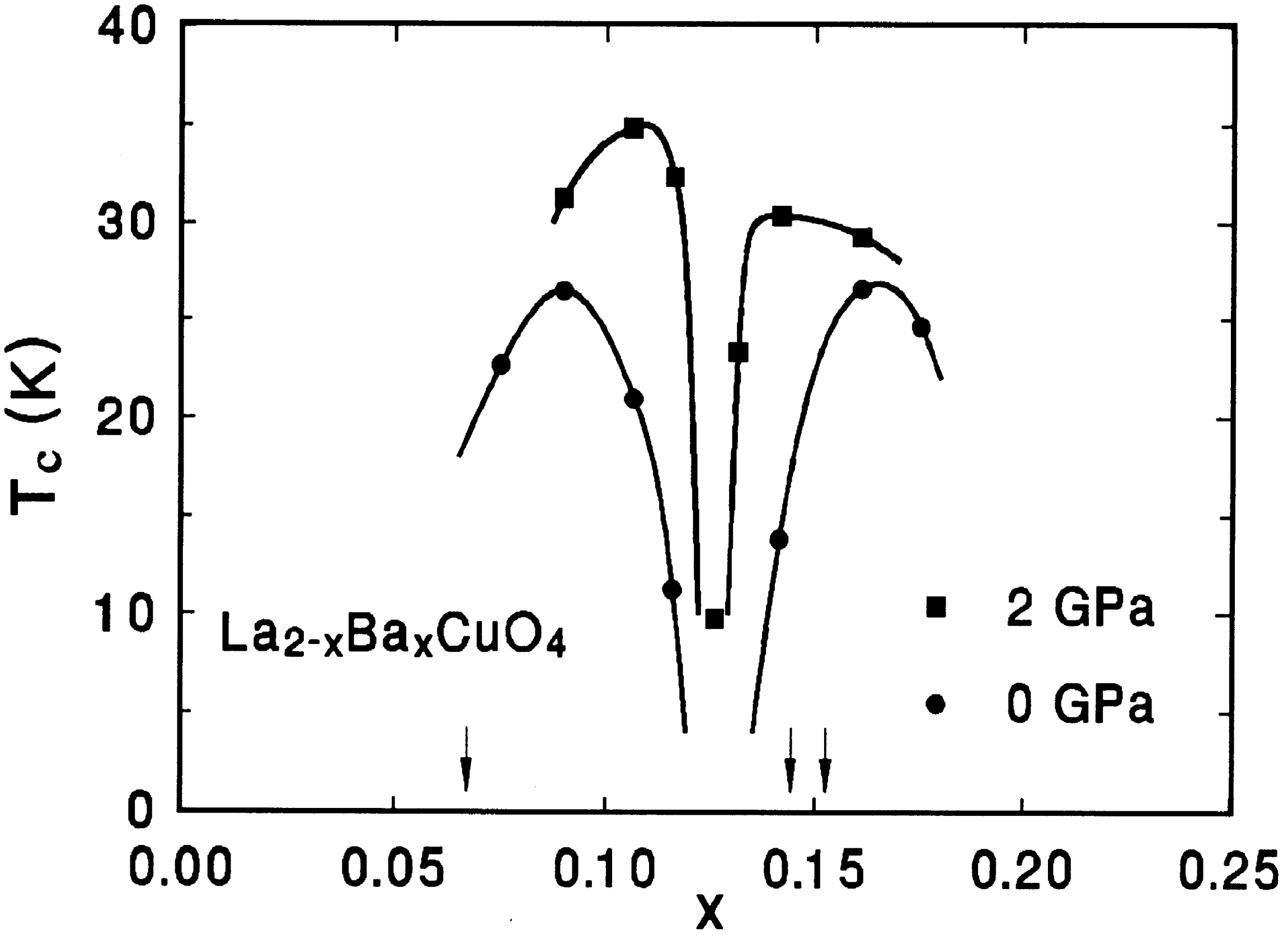

**Figure 12.** $T_c$ versus Ba content $x$ for $La_{2-x}Ba_xCuO_4$ at 0 and 2 GPa pressure from Ref. [137]. There is a marked influence of the LTT phase transition on $T_c(x)$ for $x \simeq 0.125$.

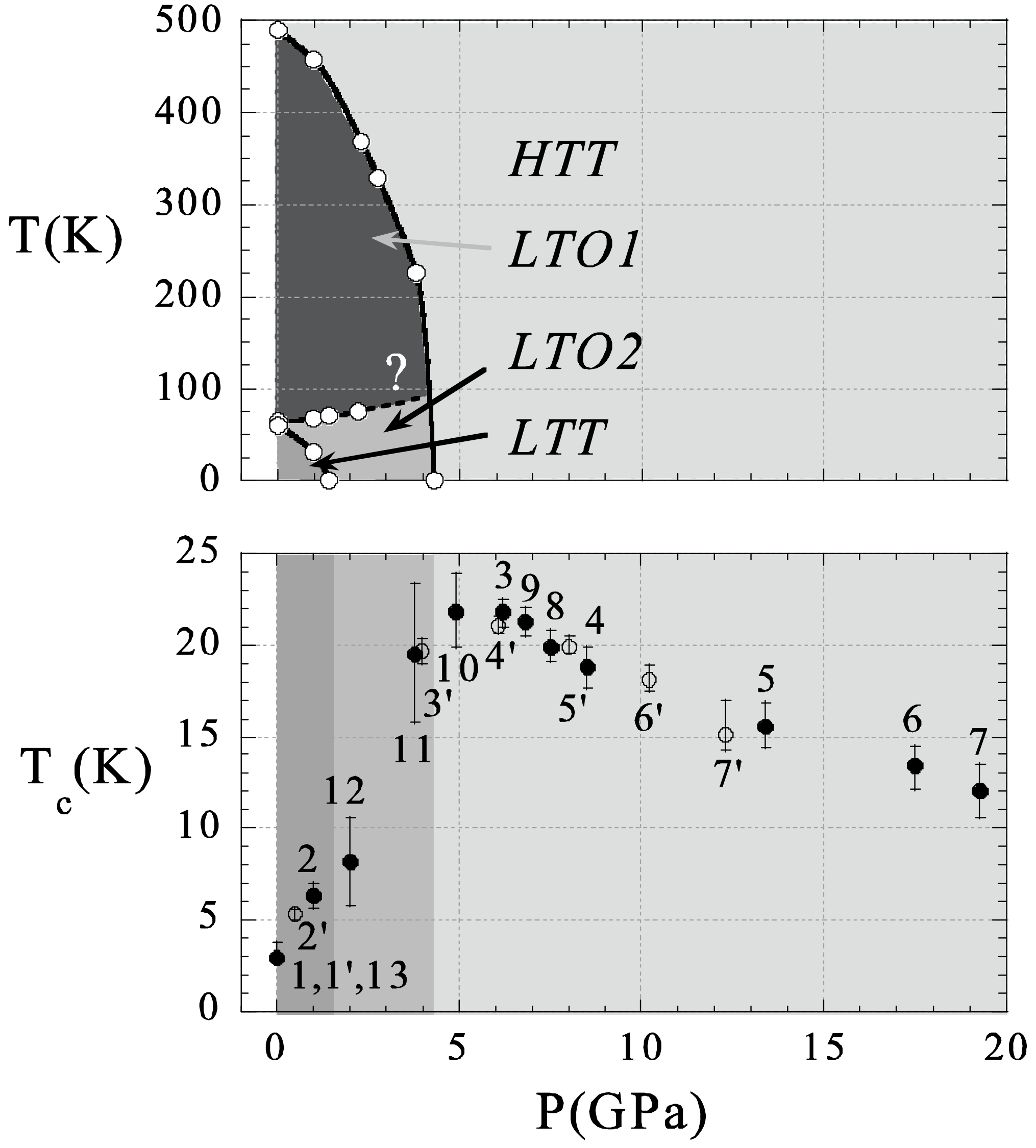

**Figure 13.** Results for $La_{1.48}Nd_{0.4}Sr_{0.12}CuO_4$ from Ref. [139]. (top) pressure-temperature phase diagram showing high-temperature tetragonal (HTT), low-temperature tetragonal (LTT), and low-temperature orthorhombic (LTO1, LTO2) phase regions. (bottom) $T_c$ versus pressure showing influence of phase transitions.

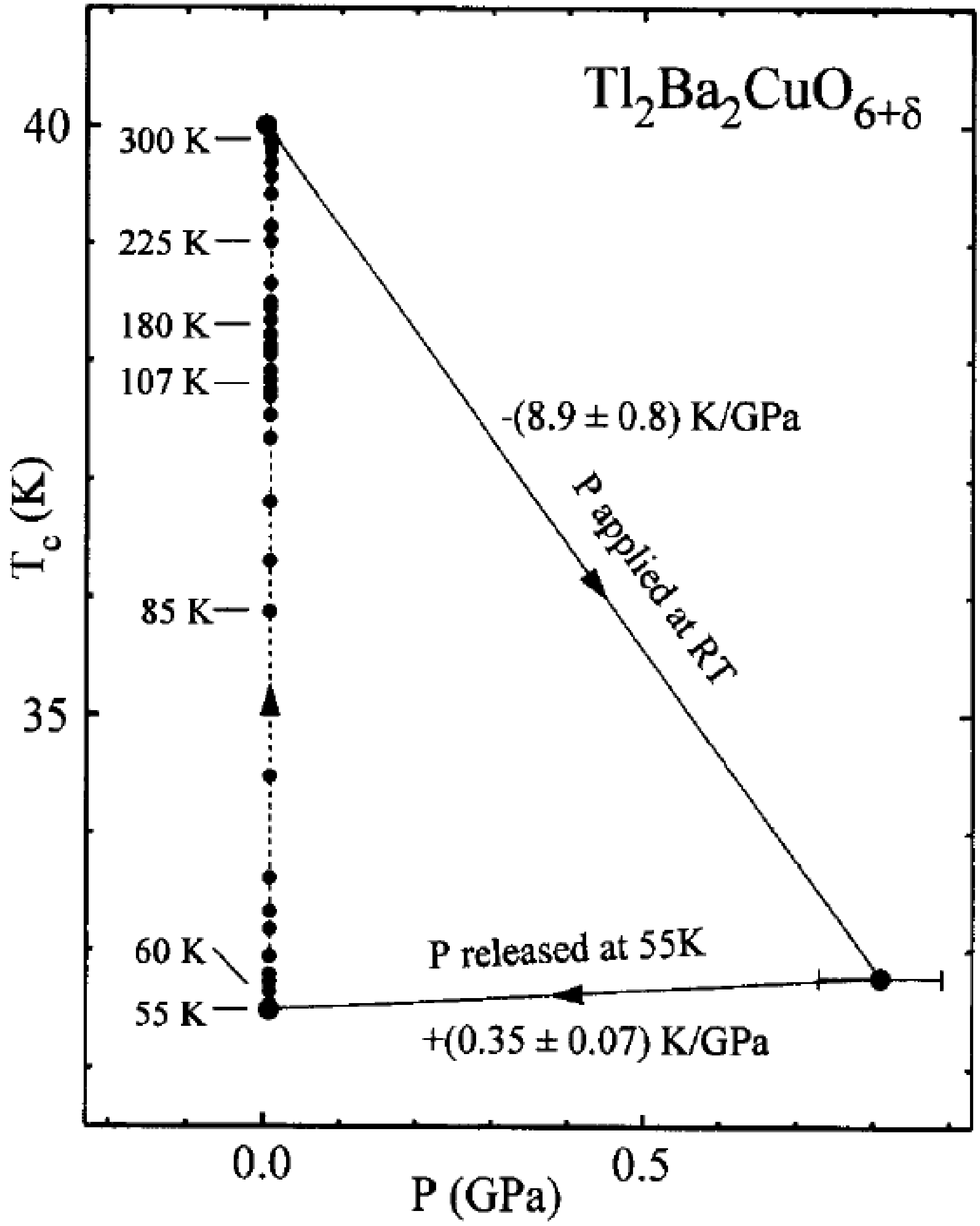

**Figure 14.** $T_c$ versus pressure for an overdoped $Tl_2Ba_2CuO_{6+\delta}$ single crystal, demonstrating the marked influence of oxygen ordering effects. Pressure is first applied at room temperature but released at 55 K, leaving the sample in a metastable state. $T_c$ relaxes back to its initial value if the sample is annealed at progressively higher temperatures to 300 K. Figure from Ref. [147].

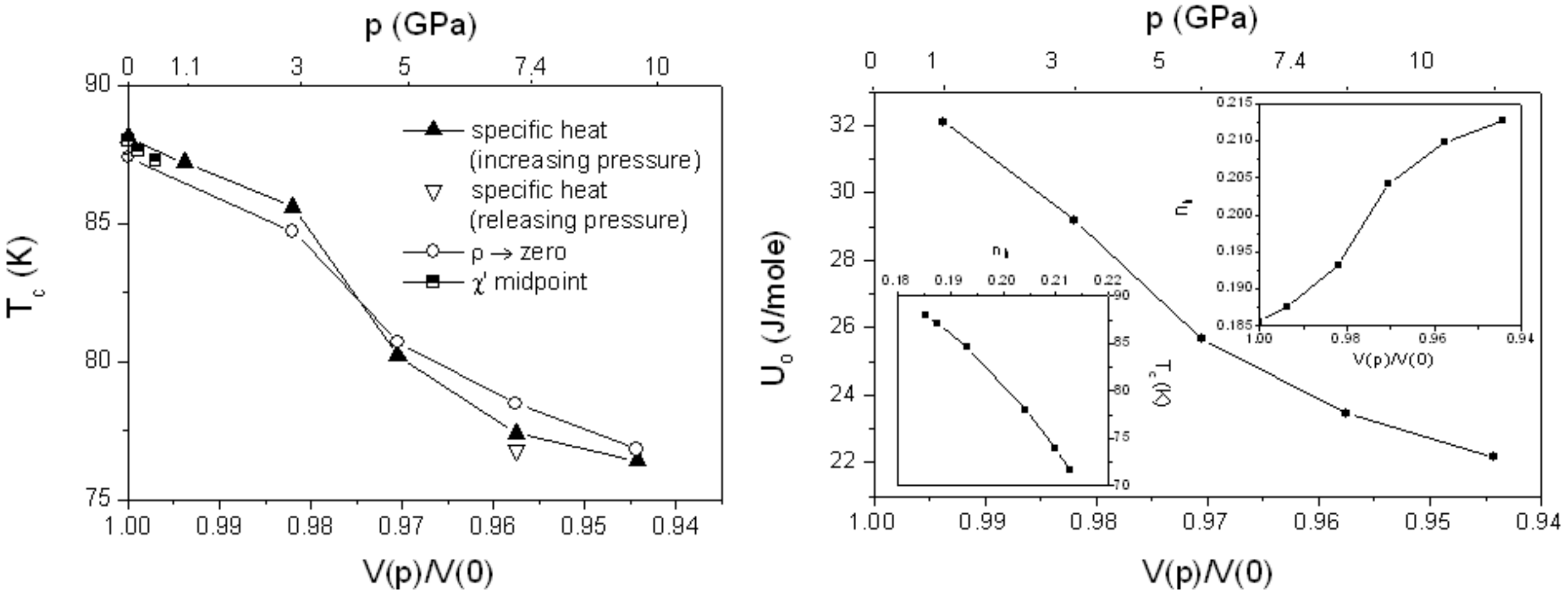

**Figure 15.** Results of specific heat measurements under pressure on an overdoped $YBa_2Cu_3O_7$ crystal from Ref. [134]. (left) $T_c$ versus both pressure P and relative volume $V(P)/V(0)$. (right) Superconducting condensation energy versus pressure $P$ and relative volume $V(P)/V(0)$. Inferred values of hole-carrier concentration $n_h$ are given in insets.

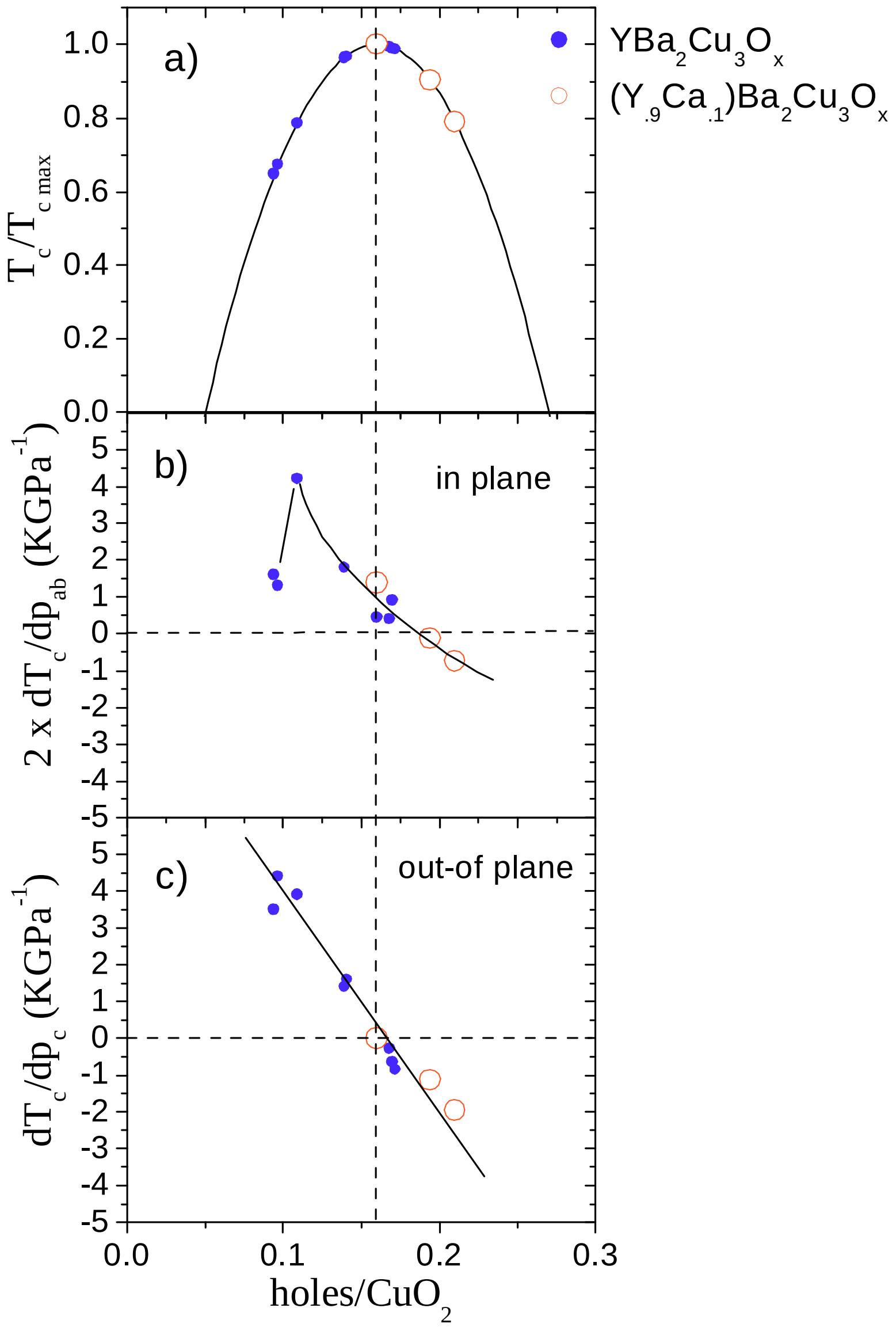

**Figure 16.** Results of ultrahigh resolution thermal expansion experiments on $Y_{1-y}Ca_yBa_2Cu_3O_x$ crystals from Ref. [127]. (a) $T_c$ versus hole concentration $n$. (b) In-plane partial pressure derivative $2(dT_c/dP_{ab})$ versus $n$. (c) Out-of-plane partial pressure derivative $dT_c/dP_c$ versus $n$.

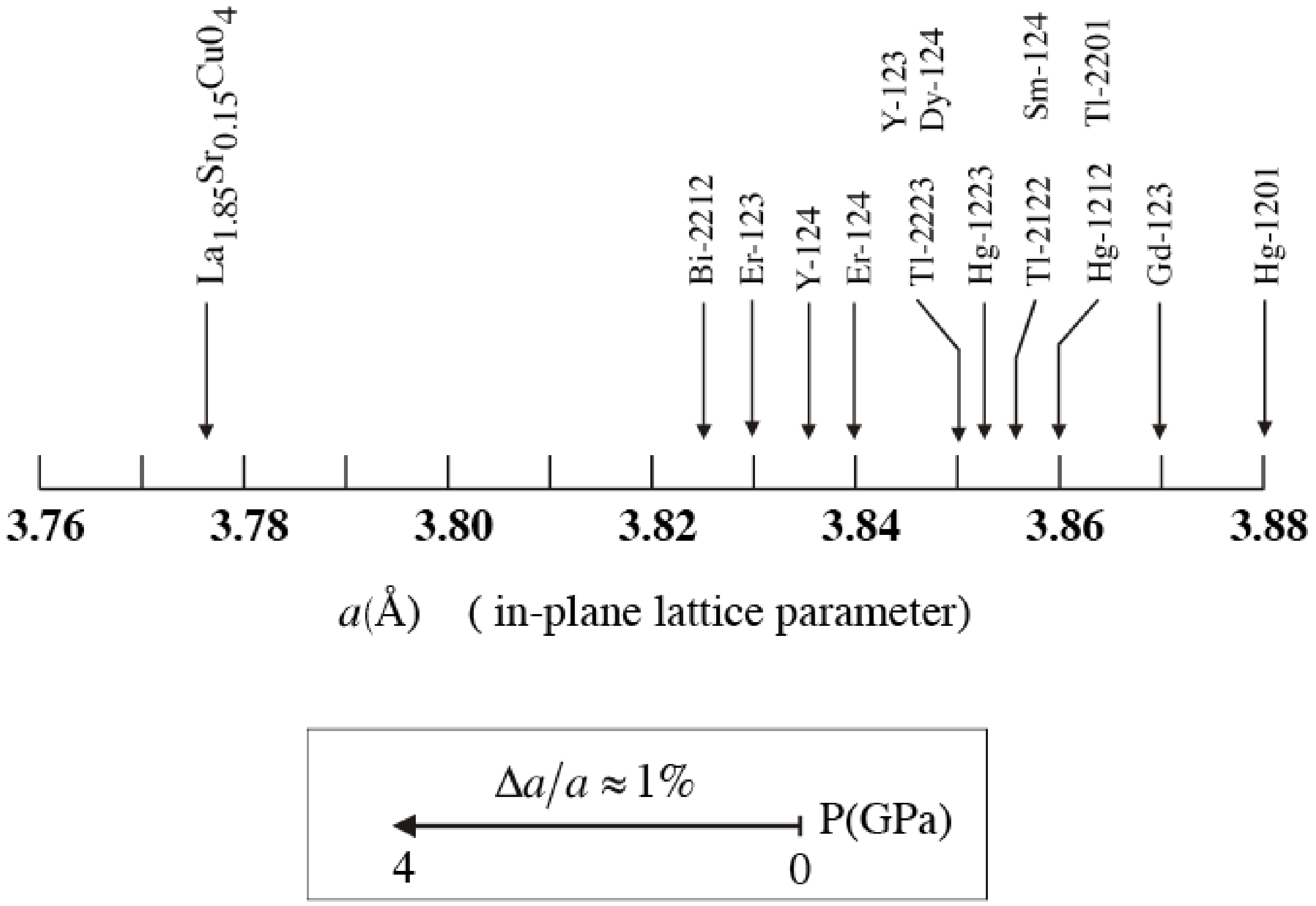

**Figure 17.** Average lattice parameter in the $CuO_2$ plane for representative HTSC systems at ambient pressure from Ref. [174].